\documentclass[journal,12pt,letterpaper,draftcls,onecolumn]{IEEEtran}%
\usepackage{amsmath}
\usepackage{amsfonts}
\usepackage{cite}
\usepackage{amssymb}
\usepackage{graphicx}
\usepackage[usenames,dvipsnames]{color}
\usepackage{epstopdf}
\usepackage[all]{xy}
\usepackage[]{mcode}

\usepackage{subcaption}
\providecommand{\U}[1]{\protect\rule{.1in}{.1in}}
\pdfoutput=1
\providecommand{\U}[1]{\protect\rule{.1in}{.1in}}

\newtheorem{lemma}{Lemma}
\newtheorem{definition}{Definition}

\newcommand{\vc}[3]{\overset{#2}{\underset{#3}{#1}}}

\newcommand{\qed}{\nobreak \ifvmode \relax \else
      \ifdim\lastskip<1.5em \hskip-\lastskip
      \hskip1.5em plus0em minus0.5em \fi \nobreak
      \vrule height0.75em width0.5em depth0.25em\fi}

\begin{document}

\title{Multicarrier Relay Selection for Full-Duplex Relay-Assisted OFDM D2D Systems}
\author{Shuping Dang, \IEEEmembership{Student Member, IEEE}, Gaojie Chen, \IEEEmembership{Member, IEEE},\\ and Justin P. Coon, \IEEEmembership{Senior Member, IEEE}
  \thanks{
  This work was supported by the SEN grant (EPSRC grant number EP/N002350/1) and the grant from China Scholarship Council  (No. 201508060323).

   The authors are with the Department of Engineering Science, University of Oxford, Parks Road, Oxford, U.K., OX1 3PJ; tel: +44 (0)1865 283 393, (e-mail: \{shuping.dang, gaojie.chen, justin.coon\}@eng.ox.ac.uk).}}

\maketitle

\begin{abstract}
In this paper, we propose a full-duplex orthogonal frequency-division multiplexing (OFDM) device-to-device (D2D) system in two-hop networks, where multiple full-duplex decode-and-forward (DF) relays assist the transmission from D2D user equipment (DUE) transmitter to DUE receiver. By such a transmission mechanism, the signal transmitted by the DUE transmitter does not need to go through a base station (BS). Meanwhile, due to the adoption of underlay D2D communication protocol, power control mechanisms are thereby necessary to be applied to mitigate the interference to conventional cellular communications. Based on these considerations, we analyze the outage performance of the proposed system, and derive the exact expressions of outage probabilities when bulk and per-subcarrier relay selection criteria are applied. Furthermore, closed-form expressions of outage probabilities are also obtained for special cases when the instantaneous channel state information (CSI) between BS and cellular user equipments (CUEs) is not accessible, so that a static power control mechanism is applied.  Subsequently, we also investigate the outage performance optimization problem by coordinating transmit power among DUE transmitter and relays, and provide a suboptimal solution, which is capable of improving the outage performance. All analysis is substantiated by numerical results provided by Monte Carlo simulations. The analytical and numerical results demonstrated in this paper can provide an insight into the full-duplex relay-assisted OFDM D2D systems, and serve as a guideline for its implementation in next generation networks.
\end{abstract}

\begin{IEEEkeywords}
Device-to-device (D2D) communications, relay selection, full-duplex system, OFDM, outage performance.
\end{IEEEkeywords}

\section{Introduction}
\IEEEPARstart{W}{ith} a rapidly increasing demand of communication services in recent years, existing communication technologies relying on infrastructure, e.g. access point (AP) and base station (BS), will soon be insufficient to meet the requirements of ubiquitous communications in the near future \cite{6815897}. As a result, device-to-device (D2D) communication has attracted a considerable amount of attention in recent years and been regarded as a promising technology for next generation networks due to its high power efficiency, high spectral efficiency and low transmission delay \cite{6805125,7217785,7163505}. D2D communication enables the direct wireless transmission between two devices (a.k.a. D2D user equipments (DUEs)) in proximity, without going through a BS. Such a flexible transmission protocol releases the design requirements of infrastructure and thereby saves transmission overheads caused by centralized coordination and management \cite{6805125}. Meanwhile, D2D communications can be classified in two categories, depending on whether frequency resources are shared between D2D communications and  traditional cellular communications, which are termed \textit{underlay} and \textit{overlay} D2D communications, respectively \cite{6805125}. It has been proved that underlay D2D communications would be able to provide a high spectrum efficiency and suit the spectrum sharing nature in next generation networks \cite{6909030,7542601,7842024}. However, the underlay D2D transmission will break up the orthogonality between D2D communications and traditional cellular communications, and the corresponding interference shall be coordinated accordingly. 

On the other hand, conventional D2D communications requiring a strong direct link between DUEs might not always be feasible in practice, as the direct link could be in deep fading and shadowing due to the existence of physical obstacles. In this scenario, D2D communications will become impractical or require a huge amount of transmit power, which will result in severe interference to cellular communications and significantly degrade the overall system performance \cite{7021996}. To solve this problem and enhance the applicability of D2D communications, relay-assisted D2D communication was proposed with decode-and-forward (DF) relays and amplify-and-forward (AF) relays in \cite{6362495} and \cite{7021996}, respectively. However, there is no exact analytical results provided in these two pioneering works. Then, the power control strategy and energy-related issues for relay-assisted D2D communications were numerically studied in \cite{7341217} and \cite{7386667}. Moreover, some practical aspects of relay-assisted D2D communication systems, e.g. transmission capacity and delay performance were investigated in \cite{7592433} and \cite{7878672}. 

To further enhance the performance of relay-assisted D2D systems, recent research also focuses on the employment of full-duplex relays, as it would double the transmission rate, as long as the self-interference (SI) can be dealt with appropriately \cite{6963403,7781718,7577711}. In \cite{6963403}, the authors proposed a novel underlay D2D communication scheme, which dynamically assigns DUE transmitters as full-duplex relays to assist cellular downlink transmissions. In \cite{7781718}, the coverage probability is analyzed for the D2D communication scenario, in which CUEs are assisted by full-duplex relays. A simple case of a pair of DUEs assisted by only one full-duplex relay is discussed in \cite{7577711}. However, the aforementioned works have not considered the application of multicarrier paradigms, which degrades their practicability in next generation networks \cite{6824752}. At the time of writing, the only two works incorporating D2D systems and multicarrier paradigms are given in  \cite{7725969,7809043}. However, these works only employ optimization techniques to provide numerical results without giving much insight into the multicarrier D2D system \textit{per se}.

Therefore, to fill the gap between relay-assisted D2D communications and multicarrier paradigms and provide a comprehensive analysis, we propose a full-duplex orthogonal frequency-division multiplexing (OFDM) D2D system assisted by multiple relays and analyze its outage performance in this paper. To be specific, DF forwarding protocols with bulk and per-subcarrier relay selections are taken into consideration, which make the proposed system more realistic for practical scenarios. To summarize, the contributions of this paper are listed infra:
\begin{enumerate}
 \item We propose a more practical system model combining relay-assisted D2D communications, OFDM systems, full-duplex transmissions and multicarrier relay selections, which suits the nature of next generation networks.
 \item We analyze the outage performance of the proposed system with multiple DF relays applying two different relay selection schemes.
 \item We derive the exact expressions of outage probabilities for all scenarios as well as the closed-form expressions for some special cases and numerically verify them.
 \item We formulate an optimization problem for the outage performance and propose suboptimal solutions to efficiently yield a better outage performance. 
 \end{enumerate} 

The rest of this paper is organized as follows. We present the system model in Section \ref{sm}. Then, outage performance for different relay selection schemes is analyzed in Section \ref{opa}. After that, we formulate the outage performance optimization problem and provide suboptimal solutions in Section \ref{opo}. Subsequently, all analysis is numerically verified by Monte Carlo simulations in Section \ref{nr}. Finally, the paper is concluded in Section \ref{c}.

\section{System Model}\label{sm}

\begin{figure}[!t]
\centering
\includegraphics[width=5.0in]{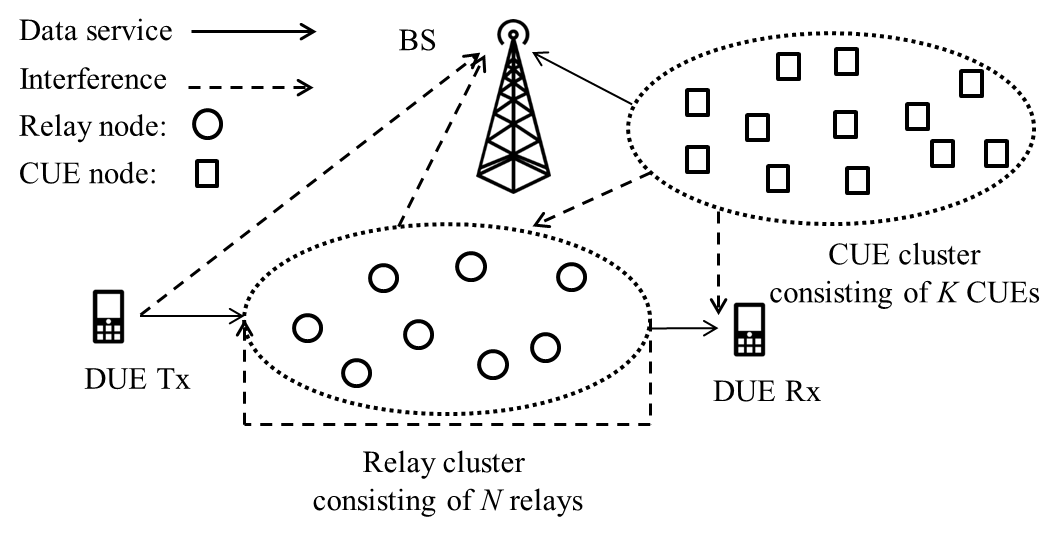}
\caption{Network model for the proposed full-duplex relay-assisted D2D system, containing one BS, $K$ CUEs in a CUE cluster, $N$ relays in a relay cluster, one DUE transmitter (source) and one DUE receiver (destination).}
\label{sys}
\end{figure}

\subsection{System framework}
The framework of the proposed system is presented in Fig. \ref{sys}, where one BS,  a pair of DUE transmitter and receiver, a cluster of $N$ full-duplex DF relays and a cluster of $K$ cellular user equipments (CUEs) are considered. Their shorthand notations are $B$, $S$, $D$, $R_{n}$ and $C_{k}$, respectively, $\forall~n\in\mathcal{N}_R=\{1,2,\dots,N\}$ and $\forall~k\in\mathcal{N}_C=\{1,2,\dots,K\}$. Meanwhile, by employing OFDM, it is supposed that there exist $K$ independent subcarriers allocated to $K$ CUEs and used by CUEs and DUEs in an underlay manner. The set of these $K$ subcarriers is denoted as $\mathcal{K}=\{1,2,\dots,K\}$. In other words, there is a unique injective mapping relation between $\mathcal{N}_C$ and $\mathcal{K}$, in order to mitigate the interference among CUEs and optimize the multiplexing gain of the cellular network\footnote{Here, we omit the subcarrier allocation process for CUEs, and assume it to be a \textit{fait accompli} as a system configuration in this paper. Details of OFDM subcarrier allocation for multiuser scenarios can be found at \cite{793310,7378853,7454319}.}.

From Fig. \ref{sys}, it is obvious that the signal and interference transmissions are in an uplink scenario\footnote{Most D2D communication systems are designed to utilize uplink cellular resources, because D2D users can monitor the received power of downlink control signals to estimate the channel between the DUE transmitter and the BS \cite{6805125}. Therefore, this will help maintain the transmit power of the DUE transmitter below a threshold, so that the interference caused by D2D communication to cellular systems can be mitigated effectively \cite{5074679}. Following this common design guideline, we also assume that D2D communication utilizes uplink spectrum resources and the downlink scenario is out of the scope of this paper.}. Therefore, the existing interference can be classified into three categories: 1) interference from active CUEs to DUE receiver and selected relay(s); 2) interference from DUE transmitter and selected relay(s) to BS; 3) SI at selected relay(s) because of the adoption of the full-duplex transmission protocol. According to the basic design guidelines of D2D communication networks \cite{6805125}, the cellular communications should be ensured with priority, and the first interference is thereby inevitable in the proposed system. In order to deal with the second interference, we have to make sure that for the $k$th subcarrier, the received aggregate interference from DUE transmitter and selected relay(s) are mitigated below a certain level. Therefore, considering an interference-limited environment \cite{7355416}, a dynamic power control mechanism is applied at the DUE transmitter and selected relay(s) on the $k$th subcarrier, which can be written as:
\begin{equation}\label{psexpres}\small
P_S(k)=\min\left\lbrace\frac{\alpha P_CG_{CB}(k)}{\xi G_{SB}(k)},\bar{P}_S\right\rbrace
\end{equation}
and
\begin{equation}\label{prexpres}\small
P_{R_{n}}(k)=\min\left\lbrace\frac{(1-\alpha)P_CG_{CB}(k)}{\xi G_{R_nB}(k)},{\bar{P}_{R}}\right\rbrace,
\end{equation}
where $P_C$ is the CUE transmit power and assumed to be the same for all CUEs; $\alpha\in(0,1)$ is a preset power coordination factor, which is used to coordinate the transmit power of DUE transmitter and relays and is the same among all subcarriers and relays; $\xi$ is a preset outage threshold for cellular communications; $\bar{P}_S$ and $\bar{P}_R$ are the maximum allowed transmit power corresponding to DUE transmitter and relays on each subcarrier; $G_{ij}(k)$ denotes the channel gain for the $k$th subcarrier, given $i\neq j$ and $i,j\in\{B,S,D\}\bigcup\mathcal{N}_R\bigcup\mathcal{N}_C$\footnote{In this paper, we assume that all channels are reciprocal and therefore have $G_{ij}(k)=G_{ji}(k)$.}, and obeys the exponential distribution with the probability density function (PDF) and cumulative distribution function (CDF) given by
\begin{equation}\small
f_{G_{ij}}(g)=e^{-g/\mu_{ij}}/{\mu_{ij}}~~\Leftrightarrow ~~F_{G_{ij}}(g)=1-e^{-g/\mu_{ij}},
\end{equation}
where $\mu_{ij}$ is the average channel gain\footnote{Because of relay and CUE clusters, we can further assume that the sizes of clusters are relatively small compared to the scale of the network. As a result, we can have the uniform $\mu_{SR}$, $\mu_{RD}$, $\mu_{CR}$, $\mu_{CD}$, $\mu_{RB}$, $\mu_{CB}$ for all relays and CUEs \cite{5555992}. Following this assumption, we can integrate all CUEs using $K$ single subcarriers in the CUE cluster into a {logically intact CUE}, termed \textit{integrated CUE}, which uses multiple subcarriers. Such an equivalent processing will ease the analysis in following sections.}.

\subsection{Decode-and-forward forwarding protocol}
Because of the interference-limited environment, we can neglect the effects of additive noise at receivers and express the instantaneous signal-to-interference ratio (SIR) from DUE transmitter to the $n$th relay (i.e. the first hop) on the $k$th subcarrier by
\begin{equation}\label{ds45ad5425dsa}\small
\Gamma_{SR_n}(k)=\frac{G_{SR_n}(k)P_S(k)}{P_C G_{CR_n}(k)+\varphi_{n}(k)},
\end{equation}
where $\varphi_n(k)$ denotes the residual SI at the $n$th relay for the $k$th subcarrier, and we assume that $\varphi_n(k)$ obeys the exponential distribution with PDF and CDF written as\footnote{Admittedly, there are also works, in which Ricean distribution is employed to model the residual SI channel, since the channel can also be regarded as a line-of-sight (LOS) path \cite{6904918}. However, according to further works on SI channel modeling\cite{6353396}, the adoption of SI channel model is subject to practical situations and employed interference cancellation techniques. Therefore, without loss of generality, we choose Rayleigh distribution in this manuscript to model the SI channel and thus the channel gain is exponentially distributed. By varying the average channel gain $\bar{\varphi}$, we can easily characterize the SI cancellation capability.}
\begin{equation}\small
f_{\varphi}(g)=e^{-g/\bar{\varphi}}/{\bar{\varphi}}~~\Leftrightarrow ~~F_{\varphi}(g)=1-e^{-g/\bar{\varphi}},
\end{equation}
where $\bar{\varphi}$ is the average residual SI.

Further assuming that there does not exist a direct transmission link between DUE transmitter and receiver due to deep fading, and the DF forwarding protocol is applied at all relays, we can express the SIR from the $n$th relay to the DUE receiver (i.e. the second hop) by
\begin{equation}\label{r187duskjzhfui1}\small
\Gamma_{R_nD}(k)=\frac{G_{R_nD}(k)P_{R_n}(k)}{P_C G_{CD}(k)}.
\end{equation}
Subsequently, we can express the equivalent end-to-end instantaneous SIR for full-duplex DF relay-assisted systems by \cite{7445895}
\begin{equation}\label{dfforwardingSNR}\small
\Gamma_{SR_nD}(k)=\min\left\lbrace\Gamma_{SR_n}(k),\Gamma_{R_nD}(k)\right\rbrace.
\end{equation}

\subsection{Relay selection schemes}\label{smrsc}
\subsubsection{Bulk selection}
In this paper, we adopt the instantaneous SIR as the performance metric to perform relay selections. By bulk selection, there will be only \textit{one} relay selected out of $N$ relays. The selection criterion can be written as
\begin{equation}\label{bulkselectioncrit}\small
\mathcal{N}_{bulk}=\{\tilde{n}\}=\bigcup_{k\in\mathcal{K}}\{\tilde{n}(k)\}=\arg\max_{n\in\mathcal{N}_R}\min_{k\in\mathcal{K}}\left\lbrace\Gamma_{SR_nD}(k)\right\rbrace,
\end{equation}
where $\tilde{n}=\tilde{n}(1)=\tilde{n}(2)=\dots=\tilde{n}(K)$ is the index of the selected relay forwarding all $K$ subcarriers, and $\tilde{n}(k)$ represents the index of the selected relay forwarding the $k$th subcarrier, $\forall~k\in\mathcal{K}$.

\subsubsection{Per-subcarrier selection}
On the other hand, by per-subcarrier selection, relays are selected by each individual subcarrier in a per-subcarrier manner and the set of all selected relays can be determined by
\begin{equation}\label{psselectioncrit}\small
\begin{split}
\mathcal{N}_{ps}&=\bigcup_{k\in\mathcal{K}}\{\tilde{n}(k)\}=\bigcup_{k\in\mathcal{K}}\left\lbrace\arg\max_{n\in\mathcal{N}_R}\left\lbrace\Gamma_{SR_nD}(k)\right\rbrace\right\rbrace.
\end{split}
\end{equation}

For clarity, a pictorial illustration and more details of the bulk and per-subcarrier selections can be found in \cite{7445895}.

\subsection{Outage probability}
After relay selection, we can define the outage event of OFDM systems by \cite{6157252}
\begin{definition}\label{defoutages323x}
An outage occurs when the end-to-end SIR of any subcarrier falls below a preset outage threshold $s$.
\end{definition}
\subsubsection{Full-duplex systems}
The full-duplex transmission would potentially have an outage performance benefit, if the residual SI can be managed below a certain level by a series of SI cancellation technologies \cite{7421941}. Consequently, we adopt the full-duplex transmission protocol in this paper for relay forwarding. As a result, for full-duplex systems, we can express the outage probability after relay selection as
\begin{equation}\small\label{husanpp}
P_{out}(s)=\mathbb{P}\left\lbrace\min_{k\in\mathcal{K}}\left\lbrace \Gamma_{SR_{\tilde{n}(k)}D}(k)\right\rbrace<s\right\rbrace,
\end{equation}
where $\mathbb{P}\left\lbrace\cdot\right\rbrace$ denotes the probability of the event enclosed.

\subsubsection{Half-duplex systems}
As an important comparison benchmark of full-duplex systems, we also give the outage probability for half-duplex systems as follows \cite{7577711}:
\begin{equation}\small\label{husanpp2}
P_{out}^{half}(s)=\mathbb{P}\left\lbrace\min_{k\in\mathcal{K}}\left\lbrace\Gamma_{SR_{\tilde{n}(k)}D}^{half}(k)\right\rbrace<s(s+2)\right\rbrace,
\end{equation}
where $\Gamma_{SR_{\tilde{n}(k)}D}^{half}(k)=\Gamma_{SR_{\tilde{n}(k)}D}(k)\vert_{\varphi_{\tilde{n}(k)}=0}$.

\section{Outage Performance Analysis}\label{opa}
By observing (\ref{psexpres}), (\ref{prexpres}) and (\ref{dfforwardingSNR}), we can find that $G_{CB}(k)$ is a common term in the first and second hops and will result in a correlation between two hops. Moreover, $G_{CD}(k)$ and $G_{SB}(k)$ will lead to correlations among relays when performing relay selections. In order to carry out analysis without considering these correlation terms, we can temporarily let them be fixed values, say $G_{CB}(k)=\bar{g}(k)$, $G_{CD}(k)=\bar{h}(k)$ and $G_{SB}(k)=\bar{l}(k)$. As a consequence, the SIRs corresponding to different relays and subcarriers can be regarded as independent. Now we can define the conditional \textit{a priori} outage probability, (i.e., the outage probability not conditioned on any form of relay selection having taken place) for the $k$th subcarrier forwarded by an arbitrary relay in the first and second hops as
\begin{equation}\small
\Xi_1\left(k\vert\bar{g}(k),\bar{l}(k) \right)=\mathbb{P}\left\lbrace\Gamma_{SR_n}(k)<s\vert\bar{g}(k),\bar{l}(k)\right\rbrace,
\end{equation}
and
\begin{equation}\label{sd54a65s4d52xxx}\small
\Xi_2\left(k\vert\bar{g}(k),\bar{h}(k) \right)=\mathbb{P}\left\lbrace\Gamma_{R_nD}(k)<s\vert\bar{g}(k),\bar{h}(k)\right\rbrace.
\end{equation}
Consequently, due to the bottleneck effect of two-hop DF relay networks (c.f. (\ref{dfforwardingSNR})), the conditional end-to-end \textit{a priori} outage probability can be determined by
\begin{equation}\label{ds5a4d65245tus}\small
\begin{split}
\Xi(k\vert\bar{g}(k),\bar{h}(k),\bar{l}(k))&=\Xi_1(k\vert\bar{g}(k),\bar{l}(k))+\Xi_2(k\vert\bar{g}(k),\bar{h}(k))-\Xi_1(k\vert\bar{g}(k),\bar{l}(k))\Xi_2(k\vert\bar{g}(k),\bar{h}(k)).
\end{split}
\end{equation}
To carry out further analysis, we should now focus on the derivations of $\Xi_1\left(k\vert\bar{g}(k),\bar{l}(k) \right)$ and $\Xi_2\left(k\vert\bar{g}(k),\bar{h}(k) \right)$. Because we have temporarily fixed $G_{CB}(k)=\bar{g}(k)$ and $G_{SB}(k)=\bar{l}(k)$, $P_S(k)$ can be viewed as a fixed coefficient, instead of a random variable (c.f. (\ref{psexpres})). Therefore, to derive $\Xi(k\vert\bar{g}(k),\bar{h}(k),\bar{l}(k))$, we first need to determine the distribution of $Z(k)=P_CG_{CR_n}(k)+\varphi_n(k)$, which can be written as
\begin{equation}\label{sdhkjash2yymens}\small
F_{Z}(z)=\begin{cases}\frac{\bar{\varphi}\left(1-e^{-\frac{z}{\bar{\varphi}}}\right)-P_C\mu_{CR}\left(1-e^{-\frac{z}{P_C\mu_{CR}}}\right)}{\bar{\varphi}-P_C\mu_{CR}},~\bar{\varphi}\neq P_C\mu_{CR}\\
1-\frac{z+\bar{\varphi}}{\bar{\varphi}}e^{-\frac{z}{\bar{\varphi}}},~~~~~~~~~~~~~~~~~~~~~~~~~~~\bar{\varphi}= P_C\mu_{CR}
\end{cases}
\end{equation}
Consequently, we can determine the PDF of $Z(k)$ by
\begin{equation}\small
f_{Z}(z)=\frac{\mathrm{d}F_{Z}(z)}{\mathrm{d}z}=\begin{cases}
\frac{e^{-\frac{z}{\bar{\varphi}}}-e^{-\frac{z}{P_C\mu_{CR}}}}{\bar{\varphi}-P_C\mu_{CR}},~~~~~\bar{\varphi}\neq P_C\mu_{CR}\\
\frac{z}{\bar{\varphi}^2}e^{-\frac{z}{\bar{\varphi}}},~~~~~~~~~~~~~~~~\bar{\varphi}= P_C\mu_{CR}
\end{cases}
\end{equation}
Denoting $W(k)=G_{SR_n}(k)/Z(k)$, we further determine the CDF of $W(k)$ by
\begin{equation}\label{1d3s5a246d542ds}\small
\begin{split}
F_{W}(w)&=1-\frac{\mu_{SR}^2}{(\mu_{SR}+P_C\mu_{CR}w)(\mu_{SR}+\bar{\varphi}w)}.
\end{split}
\end{equation}
Subsequently, by (\ref{ds45ad5425dsa}) and (\ref{1d3s5a246d542ds}), it is straightforward to obtain
\begin{equation}\label{tangjimar2s}\small
\Xi_1(k\vert\bar{g}(k),\bar{l}(k))=1-\frac{P_S^2(k)\mu_{SR}^2}{(P_S(k)\mu_{SR}+P_C\mu_{CR}s)(P_S(k)\mu_{SR}+\bar{\varphi}s)}.
\end{equation}
Then, by (\ref{prexpres}), (\ref{r187duskjzhfui1}) and (\ref{sd54a65s4d52xxx}), $\Xi_2(k\vert\bar{g}(k),\bar{h}(k))$ can be written and reduced to (\ref{pdsakjj2jssssssss}), where $\mathbb{E}\{\cdot\}$ denotes the expectation of the argument. Subsequently, substituting (\ref{tangjimar2s}) and (\ref{pdsakjj2jssssssss}) into (\ref{ds5a4d65245tus}) yields the expression of $\Xi(k\vert\bar{g}(k),\bar{h}(k),\bar{l}(k))$. 

\begin{figure*}[!t]
\begin{equation}\label{pdsakjj2jssssssss}\small
\begin{split}
\Xi_2(k\vert\bar{g}(k),\bar{h}(k))&=\mathbb{P}\left\lbrace G_{R_nD}(k)<\frac{P_C\bar{h}(k)s}{P_{R_n}(k)}\right\rbrace=\underset{P_{R_n}(k)}{\mathbb{E}}\left\lbrace F_{G_{RD}}\left(\frac{P_C\bar{h}(k)s}{P_{R_n}(k)}\right)\right\rbrace\\
&=1-e^{-\frac{P_C\bar{h}(k)s}{\bar{P}_R\mu_{RD}}}\left[1-e^{-\frac{(1-\alpha )P_C\bar{g}(k)}{\bar{P}_R\mu_{RB}\xi}}+\frac{(1-\alpha )\mu_{RD}\bar{g}(k)e^{-\frac{(1-\alpha )P_C\bar{g}(k)}{\bar{P}_R\mu_{RB}\xi}}}{(1-\alpha )\mu_{RD}\bar{g}(k)+\mu_{RB}\bar{h}(k)\xi s}\right]
\end{split}
\end{equation}
\hrule
\end{figure*}

\subsection{Bulk selection}
Subsequently, by order statistics and (\ref{bulkselectioncrit}), we can obtain the conditional \textit{a posteriori} outage probability for bulk selection and reduce it by  the binomial theorem to be
\begin{equation}\small
\begin{split}
P_{out}(s\vert \bar{\mathbf{g}},\bar{\mathbf{h}},\bar{\mathbf{l}})&=\left[1-\prod_{k=1}^{K}(1-\Xi(k\vert\bar{g}(k),\bar{h}(k),\bar{l}(k)))\right]^N=\sum_{n=0}^{N} \binom{N}{n}(-1)^n\prod_{k=1}^{K}(1-\Xi(k\vert\bar{g}(k),\bar{h}(k),\bar{l}(k)))^n.
\end{split}
\end{equation}
where $\bar{\mathbf{g}}=\{\bar{g}(1),\bar{g}(2),\dots,\bar{g}(K)\}$, $\bar{\mathbf{h}}=\{\bar{h}(1),\bar{h}(2),\dots,\bar{h}(K)\}$ and $\bar{\mathbf{l}}=\{\bar{l}(1),\bar{l}(2),\dots,\bar{l}(K)\}$; $\binom{\cdot}{\cdot}$ represents the binomial coefficient. To remove the conditions and obtain the final expression, we have to average $P_{out}(s\vert \bar{\mathbf{g}},\bar{\mathbf{h}},\bar{\mathbf{l}})$ over $\bar{\mathbf{g}}$, $\bar{\mathbf{h}}$ and $\bar{\mathbf{l}}$, which will result in a $3K$-fold integral and can be written as
\begin{equation}\label{changchangchaoji}\small
\begin{split}
P_{out}(s)&=\underset{\bar{\mathbf{g}},\bar{\mathbf{h}},\bar{\mathbf{l}}}{\iiint}P_{out}(s\vert \bar{\mathbf{g}},\bar{\mathbf{h}},\bar{\mathbf{l}}) f_{\bar{\mathbf{G}}}(\bar{\mathbf{g}})f_{\bar{\mathbf{H}}}(\bar{\mathbf{h}})f_{\bar{\mathbf{L}}}(\bar{\mathbf{l}})\mathrm{d}\bar{\mathbf{g}}\mathrm{d}\bar{\mathbf{h}}\mathrm{d}\bar{\mathbf{l}}=\sum_{n=0}^{N} \binom{N}{n}(-1)^n\left(\prod_{k=1}^{K}\Phi(k)\right),
\end{split}
\end{equation}
where 
\begin{equation}\small
f_{\bar{\mathbf{G}}}(\bar{\mathbf{g}})=\overset{K}{\underset{k=1}{\prod}}F_{G_{CB}}(\bar{g}(k))=\left(\frac{1}{\mu_{CB}}\right)^K\overset{K}{\underset{k=1}{\prod}}e^{-\frac{\bar{g}(k)}{\mu_{CB}}},
\end{equation}
\begin{equation}\small
f_{\bar{\mathbf{H}}}(\bar{\mathbf{h}})=\overset{K}{\underset{k=1}{\prod}}F_{G_{CD}}(\bar{h}(k))=\left(\frac{1}{\mu_{CD}}\right)^K\overset{K}{\underset{k=1}{\prod}}e^{-\frac{\bar{h}(k)}{\mu_{CD}}},
\end{equation}
and
\begin{equation}\small
f_{\bar{\mathbf{L}}}(\bar{\mathbf{l}})=\overset{K}{\underset{k=1}{\prod}}F_{G_{SB}}(\bar{l}(k))=\left(\frac{1}{\mu_{SB}}\right)^K\overset{K}{\underset{k=1}{\prod}}e^{-\frac{\bar{l}(k)}{\mu_{SB}}},
\end{equation}
denoting the joint PDFs corresponding to $\bar{\mathbf{g}}$, $\bar{\mathbf{h}}$ and $\bar{\mathbf{l}}$, respectively; $\Phi(k)$ is a triple integral, which is defined and simplified by the multinomial theorem as follows \cite{merris2003combinatorics}:
\begin{equation}\small
\begin{split}
\Phi(k)&=\underset{\bar{g},\bar{h},\bar{l}}{\iiint}(1-\Xi(k\vert\bar{g},\bar{h},\bar{l}))^n F_{G_{CB}}(\bar{g})F_{G_{CD}}(\bar{h})F_{G_{SB}}(\bar{l}) \mathrm{d}\bar{g}\mathrm{d}\bar{h}\mathrm{d}\bar{l}\\
&=\underset{\bar{g},\bar{h},\bar{l}}{\iiint}\underset{\mathsf{C}(n,4)}{\sum} \frac{n!(-1)^{n_2+n_3}}{\vc{\prod}{4}{\tau=1}n_\tau!}\Xi_1(k\vert\bar{g},\bar{l})^{n_2+n_4}\Xi_2(k\vert\bar{g},\bar{h})^{n_3+n_4} F_{G_{CB}}(\bar{g})F_{G_{CD}}(\bar{h})F_{G_{SB}}(\bar{l})  \mathrm{d}\bar{g}\mathrm{d}\bar{h}\mathrm{d}\bar{l},
\end{split}
\end{equation}
where $\mathsf{C}(n,T)=\left\lbrace n_1,n_2,\dots,n_T\vert\sum_{\tau=1}^Tn_\tau=n,~\forall~0\leq n_\tau\leq n\right\rbrace$, denoting the executive condition of the summation operation; $\bar{g}$, $\bar{h}$ and $\bar{l}$ are the shorthand notations of variables of integration $\bar{g}(k)$, $\bar{h}(k)$ and $\bar{l}(k)$, as all subcarriers are statistically equivalent. Then, we can utilize the interchangeability between summation and integration operations and the independence among $\bar{g}$, $\bar{h}$ and $\bar{l}$ to further reduce $\Phi(k)$ to
\begin{equation}\small
\begin{split}
\Phi(k)&=\underset{\mathsf{C}(n,4)}{\sum} \frac{n!(-1)^{n_2+n_3}}{\vc{\prod}{4}{\tau=1}n_\tau!}\int_{0}^{\infty}\phi_1(k,n_2+n_4)\phi_2(k,n_3+n_4)F_{G_{CB}}(\bar{g})\mathrm{d}\bar{g},
\end{split}
\end{equation}
where
\begin{equation}\small
\phi_1(k,n)=\int_{0}^{\infty}\Xi_1(k\vert\bar{g},\bar{l})^{n}F_{G_{SB}}(\bar{l})\mathrm{d}\bar{l},
\end{equation}
and
\begin{equation}\small
\phi_2(k,n)=\int_{0}^{\infty}\Xi_2(k\vert\bar{g},\bar{h})^{n}F_{G_{CD}}(\bar{h})\mathrm{d}\bar{h}.
\end{equation}
Now, let us focus on the derivations of $\phi_1(k,n_2+n_4)$ and $\phi_2(k,n_3+n_4)$. For $\phi_1(k,n_2+n_4)$, we can similarly employ the binomial theorem and obtain
\begin{equation}\small
\begin{split}
&\phi_1(k,n_2+n_4)=\sum_{p=0}^{n_2+n_4}\binom{n_2+n_4}{p}(-1)^p\vartheta(k),
\end{split}
\end{equation}
where $\vartheta(k)$ is determined in (\ref{youyigedatiao}); $\chi_u^{(p)}(\mathbf{a},b)$ is a defined function given by
\begin{figure*}[!t]
\begin{equation}\label{youyigedatiao}\small
\begin{split}
\vartheta(k)&=\int_{0}^{\infty}\left[\frac{P_S^2(k)\mu_{SR}^2}{(P_S(k)\mu_{SR}+P_C\mu_{CR}s)(P_S(k)\mu_{SR}+\bar{\varphi}s)}\right]^pF_{G_{SB}}(\bar{l})\mathrm{d}\bar{l}\\
&=\left(1-e^{-\frac{\alpha P_C\bar{g}}{\bar{P}_S\mu_{SB}\xi}}\right)\left[\frac{\bar{P}_S^2\mu_{SR}^2}{(\bar{P}_S\mu_{SR}+P_C\mu_{CR}s)(\bar{P}_S\mu_{SR}+\bar{\varphi}s)}\right]^p\\
&~~~~~~~~+\frac{1}{\mu_{SB}}\left(\frac{\alpha^2P_C\bar{g}^2\mu_{SR}^2}{\mu_{CR}\bar{\varphi}\xi^2s^2}\right)^p\chi_{\frac{\alpha P_C\bar{g}}{\bar{P}_S\xi}}^{(p)}\left(\left\lbrace\frac{\alpha \bar{g}\mu_{SR}}{\mu_{CR}\xi s},\frac{\alpha P_C\bar{g}\mu_{SR}}{\bar{\varphi}\xi s}\right\rbrace,\frac{1}{\mu_{SB}}\right)
\end{split}
\end{equation}
\hrule
\end{figure*}
\begin{equation}\small
\chi_u^{(p)}(\mathbf{a},b)=\int_{u}^{\infty}\frac{e^{-bx}}{\underset{\mathbf{a}}{\prod}(x+a_i)^p}\mathrm{d}x,
\end{equation}
where $\mathbf{a}=\{a_1,a_2,\dots,a_{N_a}\}$ denotes a set of $N_a$ positive numbers; $b$ and $u$ are positive numbers; $p$ is a nonnegative integer. When $p=0$, we can easily obtain
\begin{equation}\small
\chi_u^{(0)}(\mathbf{a},b)=e^{-bu}/b.
\end{equation}
When $p>0$, by partial fraction decomposition \cite{1451146}, we can determine the closed-form expression of $\chi_u^{(p)}(\mathbf{a},b)$ by
\begin{equation}\small
\begin{split}
\chi_u^{(p)}(\mathbf{a},b)&=\int_{u}^{\infty}\sum_{q=1}^{p}\sum_{i=1}^{N_a}\left[\frac{A(q,i)}{(x+a_i)^q}\right]e^{-bx}\mathrm{d}x\\
&=\sum_{q=1}^{p}\sum_{i=1}^{N_a}A(q,i)\int_{u}^{\infty}\frac{e^{-bx}}{(x+a_i)^q}\mathrm{d}x=\sum_{q=1}^{p}\sum_{i=1}^{N_a}A(q,i)e^{a_ib}\Gamma\left(1-q,b(a_i+u)\right),
\end{split}
\end{equation}
where $\{A(q,i)\}$ is a unique and real constant set, which can be derived by a recursive algorithm for any given $\mathbf{a}$ and $p$ \cite{kung1977fast}; $\Gamma(a,x)=\int_{x}^{\infty}t^{a-1}e^{-t}\mathrm{d}t$ is the incomplete gamma function.

For $\phi_2(k,n_3+n_4)$, we can derive its closed-form expression by applying the binomial theorem twice and exchanging the order of summation and integration, and obtain
\begin{equation}\small
\begin{split}
&\phi_2(k,n_3+n_4)=\sum_{p=0}^{n_3+n_4}\sum_{q=0}^{p}\binom{n_3+n_4}{p}\binom{p}{q}(-1)^p\left[1-e^{-\frac{(1-\alpha )P_C\bar{g}}{\bar{P}_R\mu_{RB}\xi}}\right]^{p-q}\theta(k),
\end{split}
\end{equation}
where $\theta(k)$ is defined and reduced to (\ref{changtiaotiaode}).
\begin{figure*}[!t]
\begin{equation}\label{changtiaotiaode}\small
\begin{split}
\theta(k)&=\int_{0}^{\infty}\left[\frac{(1-\alpha )\mu_{RD}\bar{g}e^{-\frac{(1-\alpha )P_C\bar{g}}{\bar{P}_R\mu_{RB}\xi}}}{(1-\alpha )\mu_{RD}\bar{g}+\mu_{RB}\bar{h}\xi s}\right]^{q} e^{-\left(\frac{pP_C s}{\bar{P}_R \mu_{RD}}+\frac{1}{\mu_{CD}}\right)\bar{h}}\mathrm{d}\bar{h}\\
&=\left[\frac{(1-\alpha )\mu_{RD}\bar{g}e^{-\frac{(1-\alpha )P_C\bar{g}}{\bar{P}_R\mu_{RB}\xi}}}{\mu_{RB}\xi s}\right]^{q}\left(\frac{p P_C s}{\bar{P}_R \mu_{RD}}+\frac{1}{\mu_{CD}}\right)^{q-1} \\
&~~~~~~~~\times e^{\frac{(1-\alpha )\mu_{RD}\bar{g}}{\mu_{RB}\xi s}\left(\frac{p P_C s}{\bar{P}_R \mu_{RD}}+\frac{1}{\mu_{CD}}\right)}\Gamma\left(1-q,\frac{(1-\alpha )\mu_{RD}\bar{g}}{\mu_{RB}\xi s}\left(\frac{p P_C s}{\bar{P}_R \mu_{RD}}+\frac{1}{\mu_{CD}}\right)\right)
\end{split}
\end{equation}
\hrule
\end{figure*}
Consequently, by substituting the single integral expression of $\Phi(k)$ into (\ref{changchangchaoji}), we can reduce $P_{out}(s)$ from a $3K$-fold integral to a summation of multiplications of a series of single integrals, which can be easily evaluated by standard numerical approaches. However, to the best of the authors' knowledge, there does not exist a closed-form expression of $P_{out}(s)$ when the dynamic power control mechanism is applied.

In addition, because of the demanding estimation of instantaneous CSI, the BS might not always be able to get access to $\bar{g}$, and therefore a static power control mechanism will be applied in this scenario. Specifically, the static power control mechanism will not take $\bar{g}$ into account, but replace it with a preset static power control factor\footnote{One should note that the static power control mechanism cannot always eliminate the deleterious effects of the interference from D2D communications to cellular communications. Although it could bring a better outage performance to D2D communications by releasing power control, this performance gain at the D2D side is at the price of the performance loss of cellular communications. As a consequence, the higher $\kappa$ is, the better the outage performance in D2D communications will be, and vice versa. In other words, this provides a performance trade-off between cellular communications and D2D communications, when both coexist in an underlay manner.} $\kappa$, which is determined by the statistical features of the network \cite{7637015}. Then, we can derive the closed-form expression of outage probability in (\ref{nammejkdhka25896}) for bulk selection.

\begin{figure*}[!t]
\begin{equation}\label{nammejkdhka25896}\small
\begin{split}
&P_{out}(s)=\sum_{n=0}^{N} \binom{N}{n}(-1)^n\prod_{k=1}^{K}\left[\underset{\mathsf{C}(n,4)}{\sum} \left(\frac{n!(-1)^{n_2+n_3}}{\vc{\prod}{4}{\tau=1}n_\tau!}\phi_1(k,n_2+n_4)\vert_{\bar{g}=\kappa}\phi_2(k,n_3+n_4)\vert_{\bar{g}=\kappa}\right)\right]
\end{split}
\end{equation}
\hrule
\end{figure*}

\subsection{Per-subcarrier selection}
Similarly, by (\ref{psselectioncrit}), we can derive the conditional \textit{a posteriori} outage probability for per-subcarrier selection to be
\begin{equation}\small
\begin{split}
P_{out}(s\vert \bar{\mathbf{g}},\bar{\mathbf{h}},\bar{\mathbf{l}})=1-\prod_{k=1}^{K}\left[1-\left(\Xi(k\vert\bar{g},\bar{h},\bar{l})\right)^N\right].
\end{split}
\end{equation}
In a similar manner as bulk selection, we remove the conditions by averaging $P_{out}(s\vert \bar{\mathbf{g}},\bar{\mathbf{h}},\bar{\mathbf{l}})$ over $\bar{\mathbf{g}}$, $\bar{\mathbf{h}}$ and $\bar{\mathbf{l}}$, which leads to a $3K$-fold integral and can be expressed as
\begin{equation}\label{zhentemechanga}\small
\begin{split}
P_{out}(s)&=\underset{\bar{\mathbf{g}},\bar{\mathbf{h}},\bar{\mathbf{l}}}{\iiint}P_{out}(s\vert \bar{\mathbf{g}},\bar{\mathbf{h}},\bar{\mathbf{l}}) f_{\bar{\mathbf{G}}}(\bar{\mathbf{g}})f_{\bar{\mathbf{H}}}(\bar{\mathbf{h}})f_{\bar{\mathbf{L}}}(\bar{\mathbf{l}})\mathrm{d}\bar{\mathbf{g}}\mathrm{d}\bar{\mathbf{h}}\mathrm{d}\bar{\mathbf{l}}=1-\prod_{k=1}^{K}\left(1-\Psi(k)\right),
\end{split}
\end{equation}
where $\Psi(k)$ is defined and can be simplified by the multinomial theorem as follows \cite{merris2003combinatorics}:
\begin{equation}\small
\begin{split}
\Psi(k)&=\underset{\bar{g},\bar{h},\bar{l}}{\iiint}(\Xi(k\vert\bar{g},\bar{h},\bar{l}))^N F_{G_{CB}}(\bar{g})F_{G_{CD}}(\bar{h})F_{G_{SB}}(\bar{l})  \mathrm{d}\bar{g}\mathrm{d}\bar{h}\mathrm{d}\bar{l}\\
&=\underset{\bar{g},\bar{h},\bar{l}}{\iiint}\sum_{\mathsf{C}(N,3)}\frac{N!(-1)^{n_3}}{\vc{\prod}{3}{\tau=3}n_\tau!} \Xi_1(k\vert\bar{g},\bar{l})^{n_1+n_3}\Xi_2(k\vert\bar{g},\bar{h})^{n_2+n_3} F_{G_{CB}}(\bar{g})F_{G_{CD}}(\bar{h})F_{G_{SB}}(\bar{l})  \mathrm{d}\bar{g}\mathrm{d}\bar{h}\mathrm{d}\bar{l}.
\end{split}
\end{equation}
Because of the interchangeability between summation and integration operations and the independence among $\bar{g}$, $\bar{h}$ and $\bar{l}$, we can simplify the triple integral in $\Psi(k)$ to a summation of a series of single integrals as
\begin{equation}\label{hanzitu2}\small
\begin{split}
\Psi(k)&=\sum_{\mathsf{C}(N,3)}\frac{N!(-1)^{n_3}}{\vc{\prod}{3}{\tau=1}n_\tau!}\int_{0}^{\infty}\phi_1(k,n_1+n_3)\phi_2(k,n_2+n_3)F_{G_{CB}}(\bar{g})\mathrm{d}\bar{g}.
\end{split}
\end{equation}
Finally, substituting (\ref{hanzitu2}) into (\ref{zhentemechanga}) yields the single integral expression of the outage probability for per-subcarrier selection when the dynamic power control mechanism is applied. Again, if the static power control mechanism is applied, we can express the closed-form expression of outage probability for per-subcarrier selection in (\ref{zuihoule}).
\begin{figure*}[!t]
\begin{equation}\label{zuihoule}\small
P_{out}(s)=1-\prod_{k=1}^{K}\left[1-\sum_{\mathsf{C}(N,3)}\left(\frac{N!(-1)^{n_3}}{\vc{\prod}{3}{\tau=1}n_\tau!}\phi_1(k,n_1+n_3)\vert_{\bar{g}=\kappa}\phi_2(k,n_2+n_3)\vert_{\bar{g}=\kappa}\right)\right]
\end{equation}
\hrule
\end{figure*}

\section{Outage Performance Optimization}\label{opo}
Because of the joint power control mechanism at DUE transmitter and relays, there exists a trade-off of $\alpha$ in the outage performance of relay-assisted D2D communications, which is associated with channel statistics. That is to say, there exists an optimal $\alpha^*\in(0,1)$, which is capable of minimizing the outage probability. Following this thought, we can then formulate an outage performance optimization problem infra
\begin{equation}\label{diyigezuhewenti}\small
\begin{split}
&\min_{\alpha} \{P_{out}(s)\}\\
&~\mathrm{s.t.}~0<\alpha<1.
\end{split}
\end{equation}
Because all subcarriers are statistically equivalent, this formulated optimization problem can be equivalently transfered to an optimization problem of maximizing the average end-to-end SIR regarding an individual subcarrier, written as
\begin{equation}\label{hengchangjuns2}\small
\begin{split}
&\max_{\alpha} \{\mathbb{E}\{\Gamma_{SR_nD}(k)\}\}\\
&~\mathrm{s.t.}~0<\alpha<1.
\end{split}
\end{equation}
The equivalence of these two optimization problems can be proved in a general manner in Appendix \ref{tilongbas}.

\subsection{Dynamic power control mechanism}
Because $\mathbb{E}\{\Gamma_{SR_nD}(k)\}$ is mathematically intractable, we must find another alternative objective function to approximate $\mathbb{E}\{\Gamma_{SR_nD}(k)\}$ and provide a suboptimal solution instead. Therefore, for the dynamic power control mechanism, we formulate an alternative optimization problem to approximate the original problem formulated in (\ref{hengchangjuns2}) by
\begin{equation}\label{qiuda23psjnde2122}\small
\begin{split}
&\max_{\alpha} \{\Omega(\alpha)\}\\
&~\mathrm{s.t.}~0<\alpha<1,
\end{split}
\end{equation}
where $\Omega(\alpha)$ is constructed by replacing all instantaneous channel gains by their averages in $\Gamma_{SR_nD}(k)$ except for $\bar{g}(k)$, and averaging over $\bar{g}(k)$; $\Omega(\alpha)$ can be explicitly expressed in (\ref{yigedatiaozi}). Then, we prove the quasi-concavity of the formulated problem in Appendix \ref{dongtaizhengm}, which enables it to be efficiently solved by standard optimization techniques (e.g. CVX in MATLAB \cite{grant2008cvx}), and a suboptimal $\alpha^\&$ can be yielded to improve the outage performance when the dynamic power control mechanism is applied.
\begin{figure*}[!t]
\begin{subequations}\label{yigedatiaozi}
\begin{equation}\small
\begin{split}
\Omega(\alpha)&=\underset{\bar{g}(k)}{\mathbb{E}}\left\lbrace\min\left\lbrace\frac{\mu_{SR}\min\left\lbrace\frac{\alpha P_C\bar{g}(k)}{\xi \mu_{SB}},\bar{P}_S\right\rbrace}{P_C \mu_{CR}+\bar{\varphi}},\frac{\mu_{RD}\min\left\lbrace\frac{(1-\alpha)P_C\bar{g}(k)}{\xi \mu_{RB}},{\bar{P}_{R}}\right\rbrace}{P_C \mu_{CD}}\right\rbrace\right\rbrace=\begin{cases}\omega_1(\alpha),~~~~\frac{\bar{P}_S\mu_{SR}}{P_C\mu_{CR}+\bar{\varphi}}>\frac{\bar{P}_R\mu_{RD}}{P_C\mu_{CD}}\\\omega_2(\alpha),~~~~\frac{\bar{P}_S\mu_{SR}}{P_C\mu_{CR}+\bar{\varphi}}\leq\frac{\bar{P}_R\mu_{RD}}{P_C\mu_{CD}}\end{cases}
\end{split}
\end{equation}
\begin{equation}\label{omijia1}\small
\omega_1(\alpha)=\begin{cases}
\frac{\alpha P_C\mu_{SR}\mu_{CB}}{\xi\mu_{SB}\left(P_C\mu_{CR}+\bar{\varphi}\right)}\left[1-e^{-\frac{\xi\bar{P}_R\mu_{SB}\mu_{RD}(P_C\mu_{CR}+\bar{\varphi})}{\alpha P_C^2\mu_{SR}\mu_{CB}\mu_{CD}}}\right],~~~~~~~~~0<\alpha<\varrho\\
\frac{(1-\alpha)P_C\mu_{RD}\mu_{CB}}{P_C\xi\mu_{RB}\mu_{CD}}\left[1-e^{-\frac{\bar{P}_R\xi\mu_{RB}}{(1-\alpha)P_C\mu_{CB}}}\right],~~~~~~~~~~~~~~~~~~~~~~\varrho\leq \alpha<1
\end{cases}
\end{equation}
\begin{equation}\label{omijia2}\small
\omega_2(\alpha)=\begin{cases}
\frac{\alpha P_C\mu_{SR}\mu_{CB}}{\xi\mu_{SB}\left(P_C\mu_{CR}+\bar{\varphi}\right)}\left[1-e^{-\frac{\bar{P}_S\xi\mu_{SB}}{\alpha P_C\mu_{CB}}}\right],~~~~~~~~~~~~~~~~~~~~~~~~~~~0<\alpha<\varrho\\
\frac{(1-\alpha)P_C\mu_{RD}\mu_{CB}}{P_C\xi\mu_{RB}\mu_{CD}}\left[1-e^{-\frac{\bar{P}_SP_C\xi\mu_{SR}\mu_{RB}\mu_{CD}}{(1-\alpha)P_C\mu_{RD}\mu_{CB}(P_C\mu_{CR}+\bar{\varphi})}}\right],~~~~\varrho\leq \alpha<1
\end{cases}
\end{equation}
\begin{equation}\small
\varrho=\frac{\mu_{SB}\mu_{RD}(P_C\mu_{CR}+\bar{\varphi})}{P_C\mu_{SR}\mu_{RB}\mu_{CD}+\mu_{SB}\mu_{RD}(P_C\mu_{CR}+\bar{\varphi})}
\end{equation}
\end{subequations}
\hrule
\end{figure*}

\subsection{Static power control mechanism}
Again, because it is difficult to analyze $\mathbb{E}\{\Gamma_{SR_nD}(k)\}$ directly, we propose an alternative optimization problem for the static power control case. This new optimization problem can be written as
\begin{equation}\label{musat}\small
\begin{split}
&\max_{\alpha}\{\gamma(\alpha)\}\\
&~\mathrm{s.t.}~0<\alpha<1,
\end{split}
\end{equation}
where $\gamma(\alpha)=\min\{\gamma_1(\alpha),\gamma_2(\alpha)\}$; $\gamma_1(\alpha)$ and $\gamma_2(\alpha)$ are given below:
\begin{equation}\small
\gamma_1(\alpha)=\frac{\mu_{SR}\min\left\lbrace\frac{\alpha P_C\kappa}{\xi \mu_{SB}},\bar{P}_S\right\rbrace}{P_C \mu_{CR}+\bar{\varphi}},
\end{equation}
and
\begin{equation}\small
\gamma_2(\alpha)=\frac{\mu_{RD}\min\left\lbrace\frac{(1-\alpha)P_C\kappa}{\xi \mu_{RB}},{\bar{P}_{R}}\right\rbrace}{P_C \mu_{CD}},
\end{equation}
which are produced by replacing all instantaneous channel gains in $\Gamma_{SR_nD}(k)$ by their averages. Subsequently, we can prove this alternative optimization problem to be quasi-concave in Appendix \ref{sajdksahdakwu}. As a result, this optimization problem can be efficiently solved by using standard optimization techniques, and a suboptimal $\alpha^\&$ can be yielded to improve the outage performance when the static power control mechanism is applied.

\begin{figure*}[!t]
    \centering
    \begin{subfigure}[t]{0.5\textwidth}
        \centering
        \includegraphics[width=3.5in]{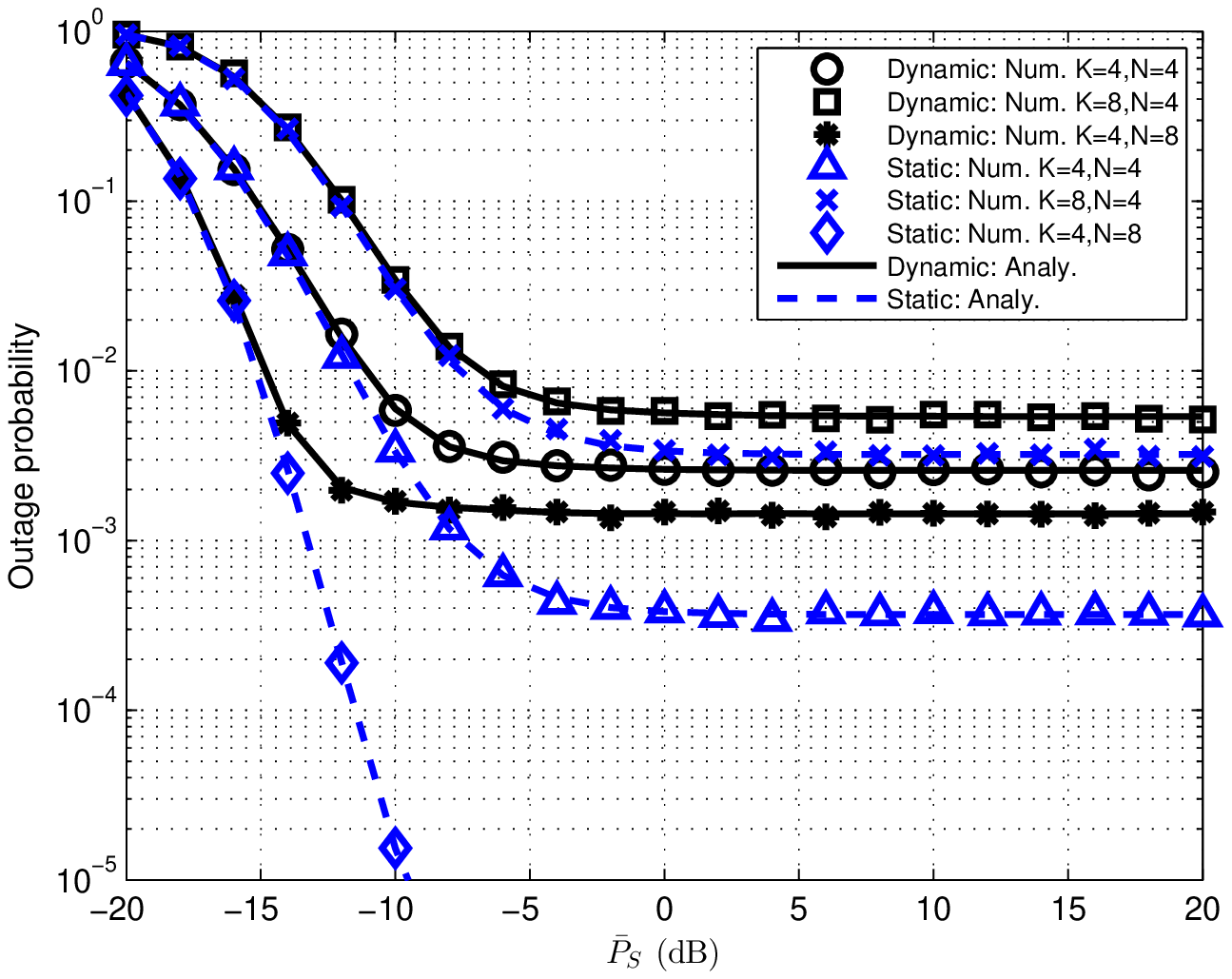}
        \caption{Bulk selection scheme}
    \end{subfigure}%
~
    \begin{subfigure}[t]{0.5\textwidth}
        \centering
        \includegraphics[width=3.5in]{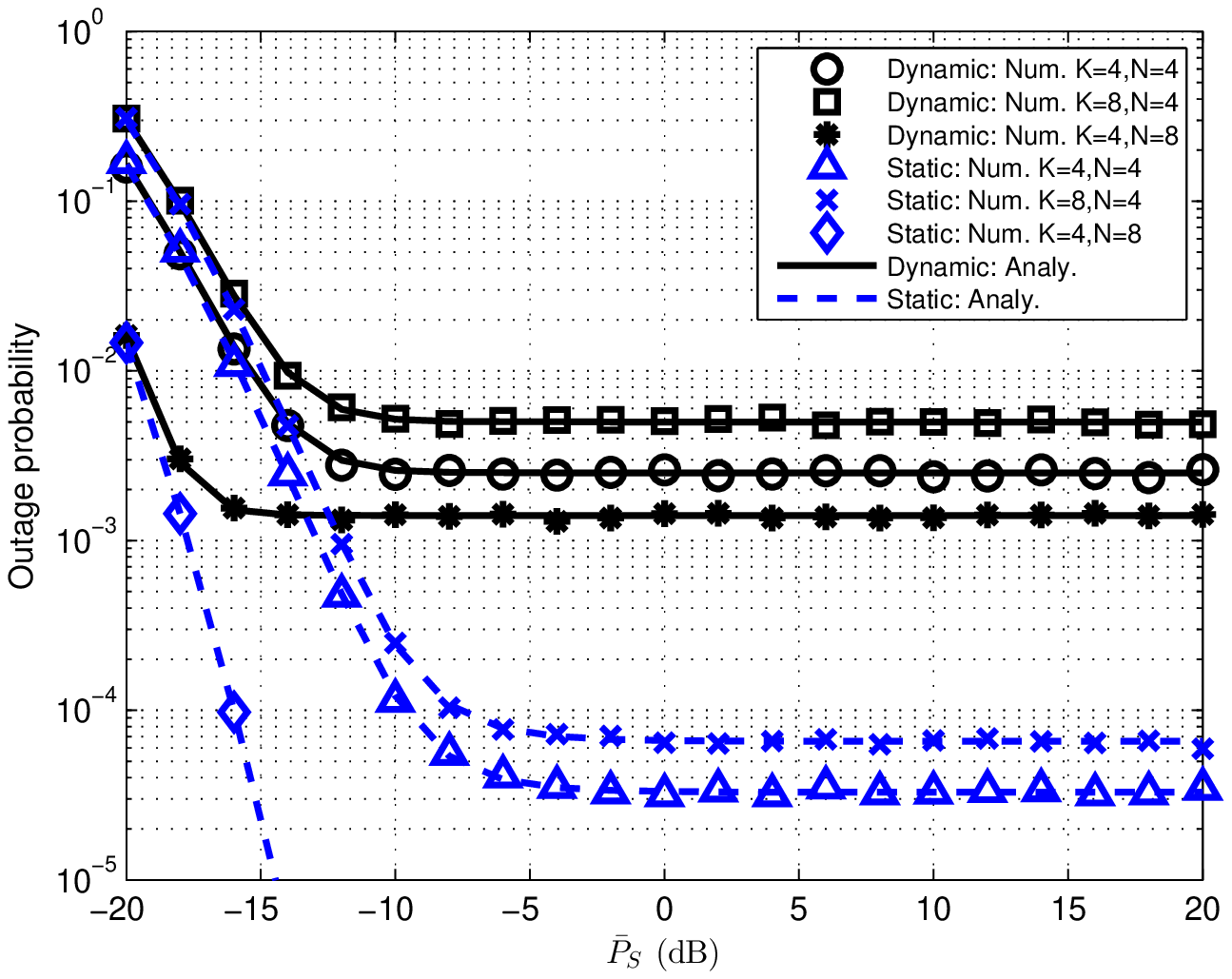}
        \caption{Per-subcarrier selection scheme}
    \end{subfigure}
    \caption{Outage probability of D2D communications vs. $\bar{P}_S$ for bulk and per-subcarrier selection schemes, given $\bar{P}_R=\bar{P}_S$.}
    \label{verification}
\end{figure*}

\section{Numerical Results}\label{nr}
\subsection{Verifications of analytical results}\label{dsajkdhkj2dsdsd}
To verify the analysis presented in Section \ref{opa} for different relay selection schemes and power control mechanisms, we carried out Monte Carlo simulations in this section. First, we set $\mu_{SR}=\mu_{RD}= 30$ dB, $\mu_{SB}=\mu_{RB}=10$ dB, $\mu_{CR}=\mu_{CD}=2$ dB, $\mu_{CB}=20$ dB and $\bar{\varphi}=5$ dB; also, we normalize $s=\xi=P_C=1$ and let $\alpha=0.5$ (without considering power coordination for the time being) and $\kappa=4$ for implementing the static power control mechanism. We further suppose $\bar{P}_R=\bar{P}_S$ and vary this transmit power limit to substantiate (\ref{changchangchaoji}), (\ref{nammejkdhka25896}), (\ref{zhentemechanga}) and (\ref{zuihoule}). The simulation results are shown in Fig. \ref{verification} for bulk and per-subcarrier selection schemes with different numbers of subcarriers and relays. 

From Fig. \ref{verification}, we can see that the theoretical results perfectly match the numerical results, which validate the correctness of our analysis given in Section \ref{opa}. In addition, the per-subcarrier selection scheme outperforms the bulk selection scheme in terms of outage probability when $\bar{P}_S$ and $\bar{P}_R$ are small. However, with the increase of $\bar{P}_S$ and $\bar{P}_R$, outage probabilities corresponding to both selection schemes get close when dynamic power control mechanism is applied, which indicates that the dynamic power control mechanism dominates the outage performance as long as $\bar{P}_S$ and $\bar{P}_R$ are sufficiently large, instead of relay selections. On the contrary, this is not the case for the static power control mechanism, as the correlation term $\bar{g}(k)$ is out of consideration, which will result in a significant performance improvement by using per-subcarrier selection.

Meanwhile, some important features of the proposed full-duplex relay-assisted OFDM D2D system can also be shown in Fig. \ref{verification}. First, the power control mechanisms constraint the improvement of outage performance by increasing $\bar{P}_S$ and $\bar{P}_R$, as it is dominated by the interference to cellular communications when $\bar{P}_S$ and $\bar{P}_R$ are large. Second, both dynamic and static power control mechanisms share a similar outage performance when $\bar{P}_S$ and $\bar{P}_R$ are small, but will have a performance gap when $\bar{P}_S$ and $\bar{P}_R$ increase. The performance gap is mainly determined by the static power control factor $\kappa$. Besides, increasing the number of subcarriers $K$ will lead to a worse outage performance, as all subcarriers have to be ensured not in outage (c.f. \textit{Definition \ref{defoutages323x}}). On the other hand, increasing the number of relays $N$ will yield a better outage performance, because the DUE transmitter can have more options when performing multicarrier relay selections. Moreover, the positive effect on outage performance by increasing $N$ for the static power control mechanism is much more obvious than that for the dynamic power control mechanism, since there is one less correlation term when the static power control mechanism is applied. These results provide guidelines for the design of OFDM D2D systems, when coexisting with traditional cellular systems in an underlay manner.

\subsection{Performance benefits by relay selections and comparisons}
\subsubsection{Effects of $\mu_{SR}$ and $\mu_{RD}$}
\begin{figure}[!t]
\centering
\includegraphics[width=5.0in]{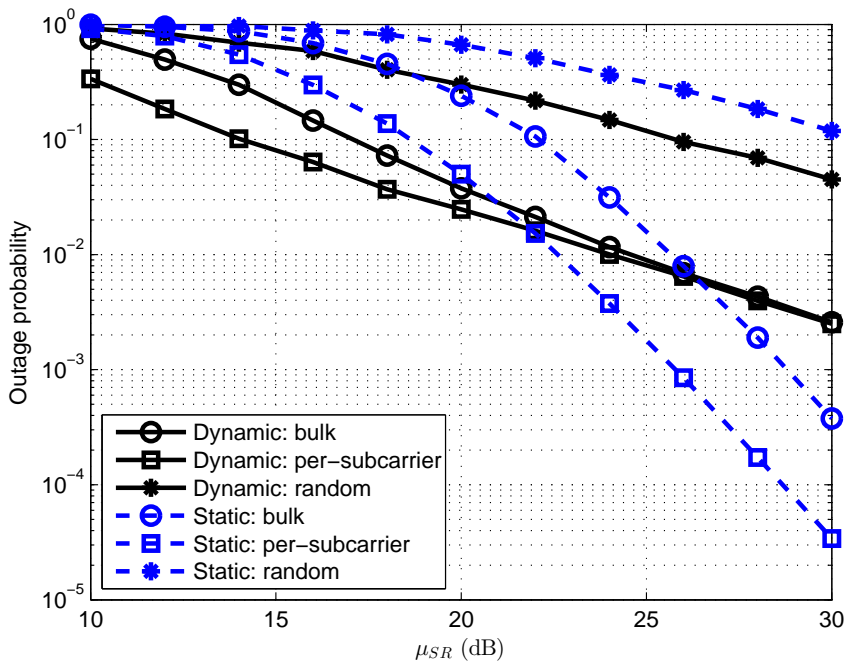}
\caption{Numerical results: outage probability of D2D communications vs. the average channel gain $\mu_{SR}$, given $\mu_{RD}=\mu_{SR}$ and $N=K=4$.}
\label{comparison_muSR_muRD}
\end{figure}

In order to study the effects of the D2D transmission links, we focus on $\mu_{SR}$ and $\mu_{RD}$ in this subsection, which are directly related to the quality of relay-assisted D2D communications. To simulate, we normalize $\bar{P}_S=\bar{P}_R=1$, set $N=K=4$, and maintain other settings as the same in the last subsection. Then, we assume $\mu_{RD}=\mu_{SR}$, and vary them to obtain the relation between the outage probability and the quality of D2D transmission links. Meanwhile, to illustrate the performance benefits brought by relay selections, we also take random relay selection as a comparison benchmark in our simulations. The numerical results are illustrated in Fig. \ref{comparison_muSR_muRD}. By these numerical results, it is clear that the outage performance yielded by bulk and per-subcarrier selections will get close with increasing $\mu_{SR}$ and $\mu_{RD}$ when the dynamic power control mechanism is applied, while this does not happen for the case of static power control mechanism. Furthermore, compared to the random selection scheme, both bulk and subcarrier selection schemes own outage performance gains, which validate the effectiveness of multicarrier relay selections in full-duplex relay-assisted OFDM D2D systems. 

\subsubsection{Effects of $\mu_{SB}$ and $\mu_{RB}$}
\begin{figure}[!t]
\centering
\includegraphics[width=5.0in]{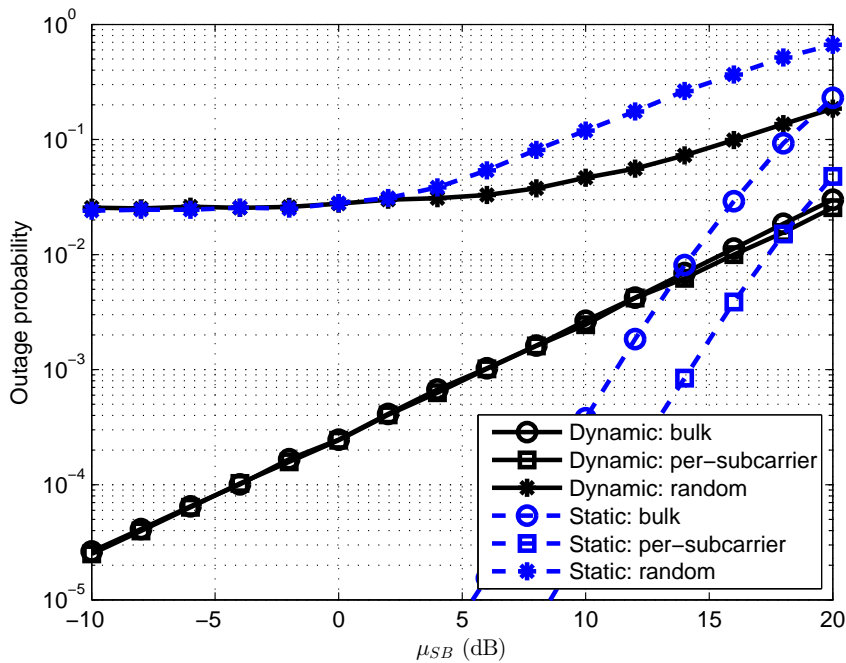}
\caption{Numerical results: outage probability of D2D communications vs. the average channel gain $\mu_{SB}$, given $\mu_{RB}=\mu_{SB}$ and $N=K=4$.}
\label{comparison_muSB_muRB}
\end{figure}

The channels between BS and DUE transmitter as well as relays are important, as they are related to the power control mechanisms. Here, we also investigate them via $\mu_{SB}$ and $\mu_{RB}$. By taking a similar simulation configurations as above, and fixing $\mu_{SR}=\mu_{RD}=30$ dB, we assume $\mu_{RB}=\mu_{SB}$, and vary them to illustrate how the qualities of these channels affect the outage performance of D2D communications. The numerical results are given in Fig. \ref{comparison_muSB_muRB}. We can find from this figure that increasing $\mu_{SB}$ and $\mu_{RB}$ will yield a significantly negative impact on the outage performance, because on average, less transmit power will be allowed for D2D communications when $\mu_{SB}$ and $\mu_{RB}$ go large. Meanwhile, because of the correlation term $\bar{l}(k)$, bulk and per-subcarrier selections have a similar outage performance when $\mu_{SR}$ and $\mu_{RD}$ are large.

\subsubsection{Effects of $\mu_{CR}$ and $\mu_{CD}$}
\begin{figure}[!t]
\centering
\includegraphics[width=5.0in]{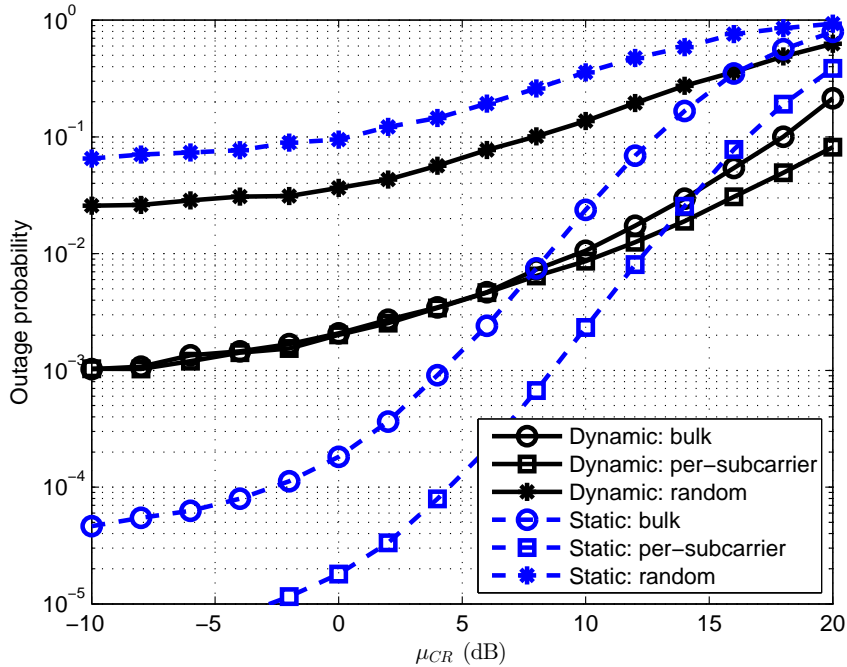}
\caption{Numerical results: outage probability of D2D communications vs. the average channel gain $\mu_{CR}$, given $\mu_{CD}=\mu_{CR}$ and $N=K=4$.}
\label{comparison_muCR_muCD}
\end{figure}

Additionally, the channels between CUE and relay as well as DUE receiver determine the quality of signal reception, as the transmitted signals from CUEs are regarded as interference to relays and the DUE receiver. Here, we let $\mu_{CD}=\mu_{CR}$, and change them to observe their impacts on the D2D signal reception. The numerical results are shown in Fig. \ref{comparison_muCR_muCD}. The numerical results demonstrated in Fig. \ref{comparison_muCR_muCD} are aligned with our expectation that increasing $\mu_{CR}$ and $\mu_{CD}$ will produce a destructive effect on outage performance, as a large interference from cellular communications will exist at relays and the DUE receiver. Meanwhile, because of the correlation term $\bar{h}(k)$, with decreasing $\mu_{CR}$ and $\mu_{CD}$, the performance curves of bulk and per-subcarrier selections get close.

\begin{figure*}[!t]
    \centering
    \begin{subfigure}[t]{0.5\textwidth}
        \centering
        \includegraphics[width=3.5in]{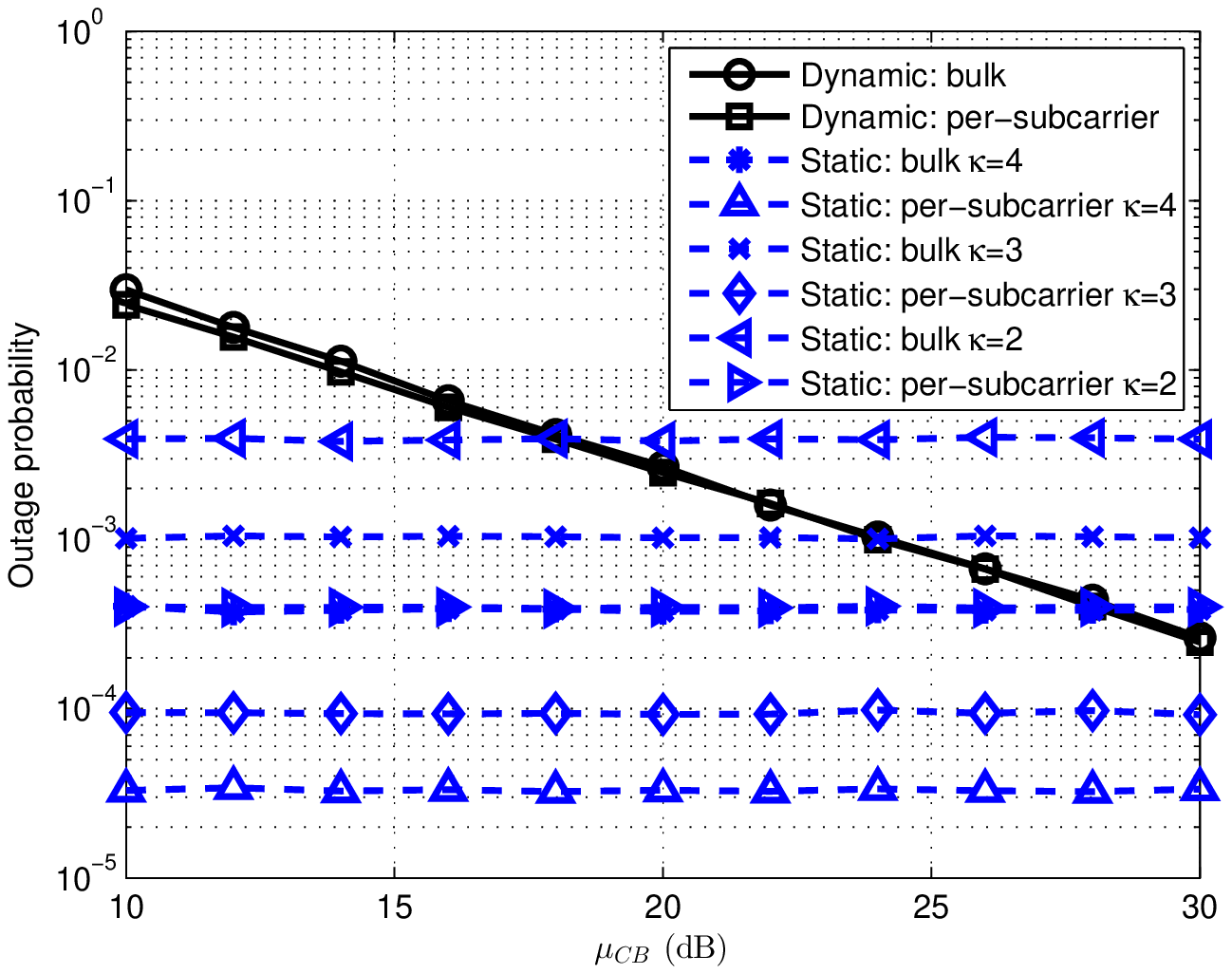}
        \caption{D2D communications}
    \end{subfigure}%
~
    \begin{subfigure}[t]{0.5\textwidth}
        \centering
        \includegraphics[width=3.5in]{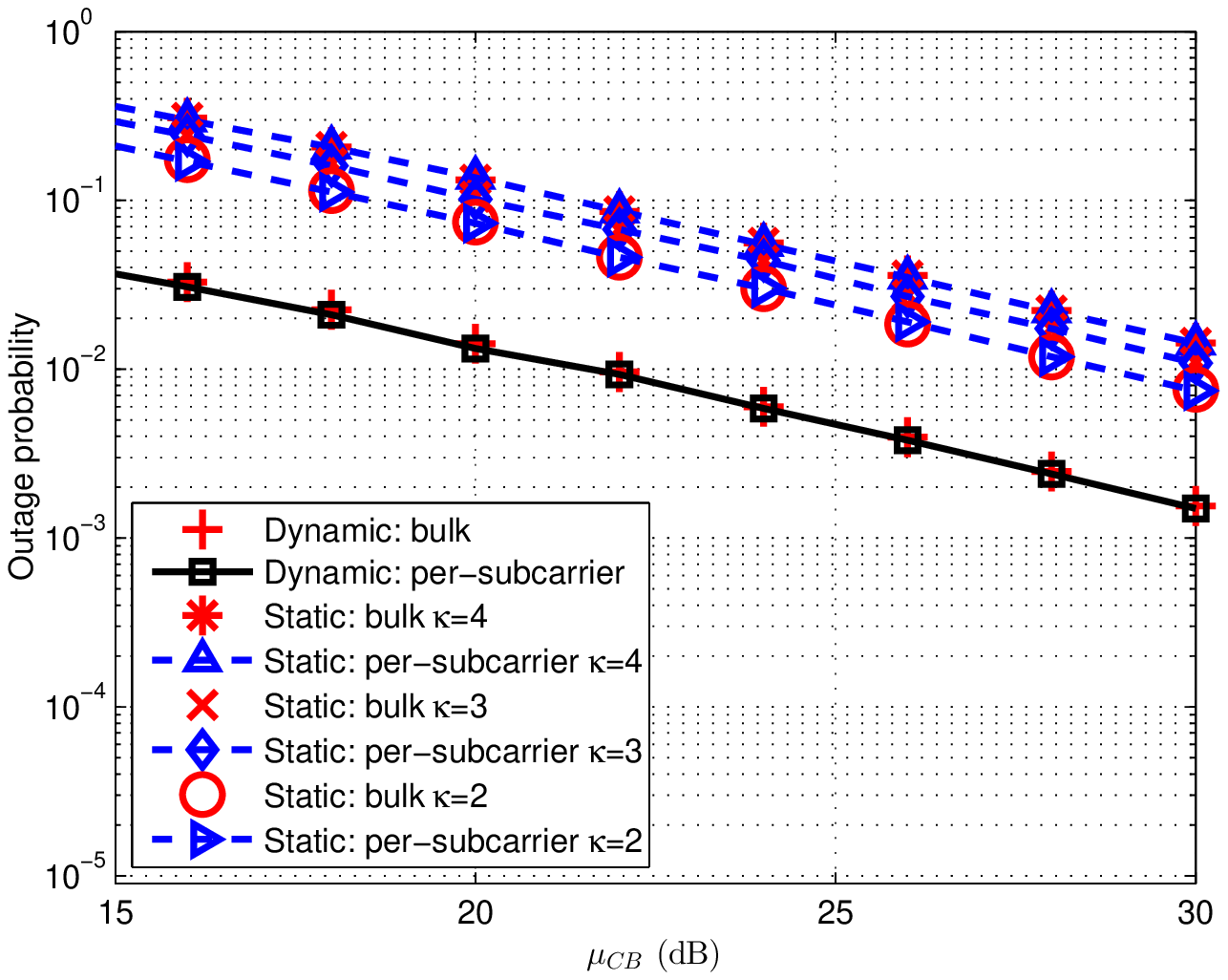}
        \caption{Cellular communications}
    \end{subfigure}
    \caption{Numerical results: outage probability vs. $\mu_{CB}$ for D2D and cellular communications, given $N=K=4$.}
    \label{muCBcomparison}
\end{figure*}

\subsection{Comparison between dynamic and static power control mechanisms}
The performance difference between dynamic and static power control mechanisms mainly depends on $\mu_{CB}$ and $\kappa$. In this subsection, we study the relation between $\mu_{CB}$ and outage performance with different $\kappa$. To provide a comprehensive analysis of the effects of $\mu_{CB}$ on both D2D and cellular communications, we should take the outage probabilities from both sides into consideration. Taking the same simulation configurations specified in Section \ref{dsajkdhkj2dsdsd} and normalizing $\bar{P}_S=\bar{P}_R=1$, we carry out the numerical simulations and present the numerical results in Fig. \ref{muCBcomparison} for both D2D and cellular communications. 

From Fig. \ref{muCBcomparison} (a), we can observe that the outage probabilities regarding bulk and per-subcarrier selections get close at high $\mu_{CB}$, which indicates that the correlation term $\bar{g}(k)$ plays a dominant role in the relay selection process, and produces a deleterious impact on the performance gain. This sub-figure also provides a hint to choose appropriate power control mechanisms. Meanwhile, when the static power control mechanism is applied, a lower $\kappa$ will lead to a better outage performance, and the performance is independent from $\mu_{CB}$, as $\bar{g}(k)$ is not taken into consideration for static power control. On the other hand, by observing Fig. \ref{muCBcomparison} (b), the numerical results verify our analysis that the potential performance gain in D2D communications brought by the static power control mechanism is at the price of the performance  loss in cellular communications. Besides, as we can also see, the relay selection schemes adopted by DUEs does not matter to the cellular communications, because on average, all relays are viewed equivalently to the BS, and so is their interference.

\begin{figure*}[!t]
    \centering
    \begin{subfigure}[t]{0.5\textwidth}
        \centering
        \includegraphics[width=3.5in]{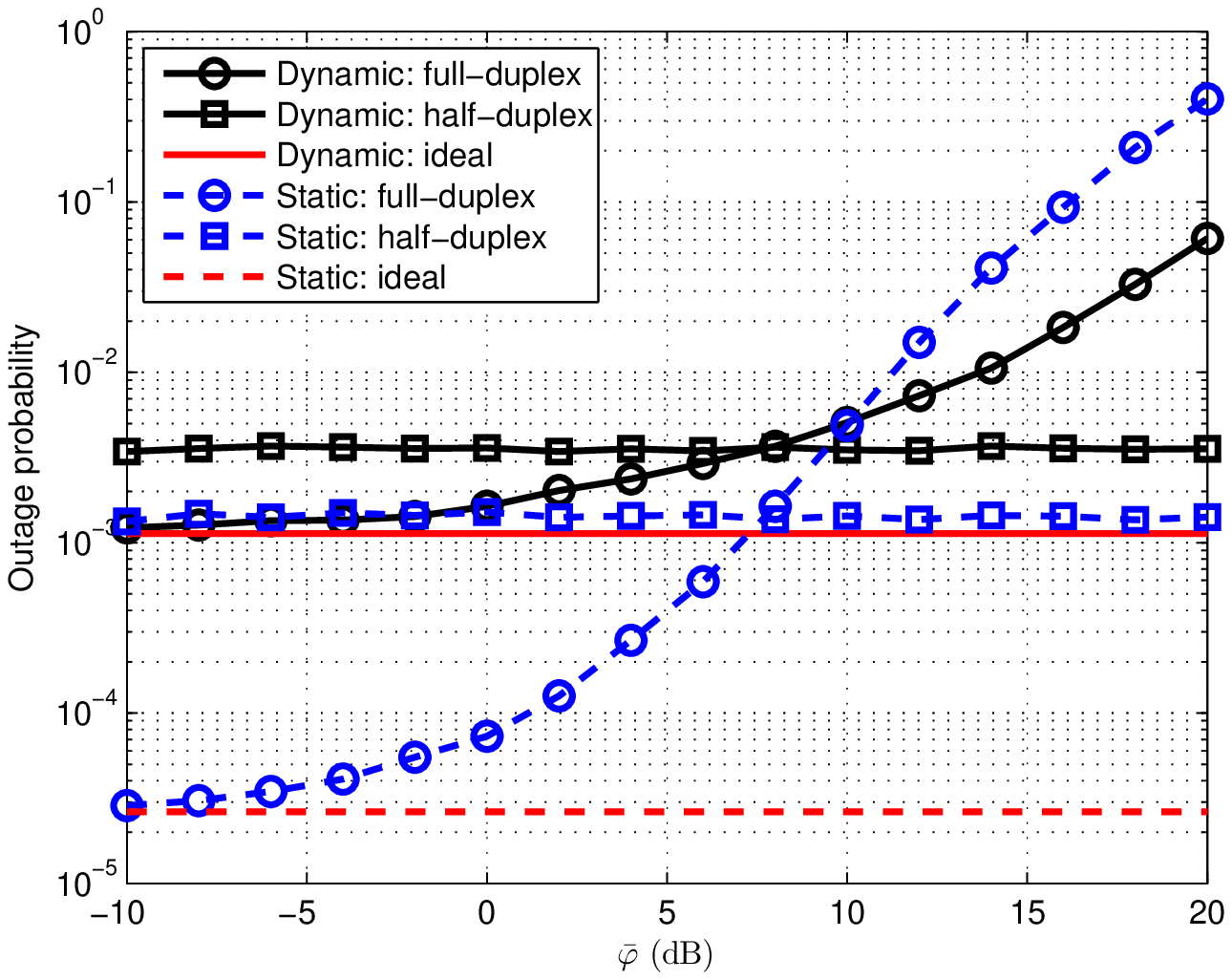}
        \caption{Bulk selection scheme}
    \end{subfigure}%
~
    \begin{subfigure}[t]{0.5\textwidth}
        \centering
        \includegraphics[width=3.5in]{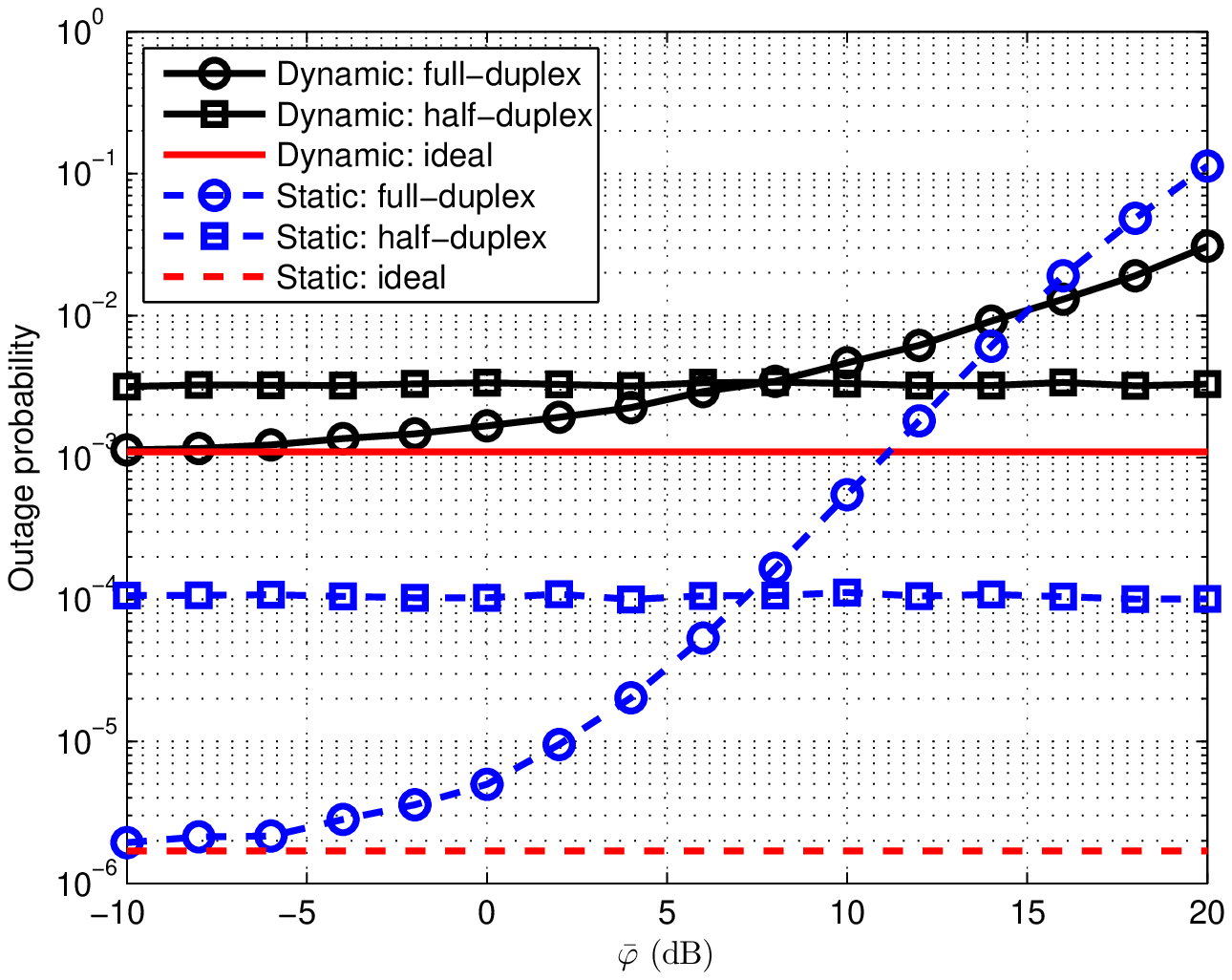}
        \caption{Per-subcarrier selection scheme}
    \end{subfigure}
    \caption{Numerical results: outage probability of D2D communications vs. $\bar{\varphi}$ for bulk and per-subcarrier selection schemes, given $N=K=4$.}
    \label{comparison_varphi}
\end{figure*}

\subsection{Comparison between full-duplex and half-duplex transmissions}
As it is well known that full-duplex transmission does not always outperform half-duplex transmission \cite{6365875}, the performance difference between them is mainly dependent on the mean of the residual SI term, i.e. $\bar{\varphi}$. Therefore, we study $\bar{\varphi}$ in this subsection and compare the outage performance provided by full-duplex and half-duplex transmissions (c.f. (\ref{husanpp}) and (\ref{husanpp2})). Meanwhile, we also provide an ideal case that the residual SI can be mitigated to a noise level and is thus negligible to show the ideal scenario of full-duplex transmission for comparison purposes. This ideal case can be produced by
\begin{equation}\small
P_{out}^{ideal}(s)=\underset{\bar{\varphi}\rightarrow 0}{\lim}P_{out}(s).
\end{equation}
Taking the same simulation configurations given in Section \ref{dsajkdhkj2dsdsd} and normalizing $\bar{P}_S=\bar{P}_R=1$, we carry out the numerical simulations and all simulation results are shown in Fig. \ref{comparison_varphi}. From this figure, it is obvious that the priority of transmission protocols depends on $\bar{\varphi}$, and full-duplex transmission does have the potential to provide a better outage performance, as long as a satisfactory SI elimination technology can be utilized to reduce $\bar{\varphi}$ below a certain level. The analytical derivation of the critical value of $\bar{\varphi}$ below which the full-duplex transmission outperforms the half-duplex transmission (i.e. the cross point of two outage probability curves) would be worth investigating as a future work.

\begin{figure*}[!t]
    \centering
    \begin{subfigure}[t]{0.5\textwidth}
        \centering
        \includegraphics[width=3.5in]{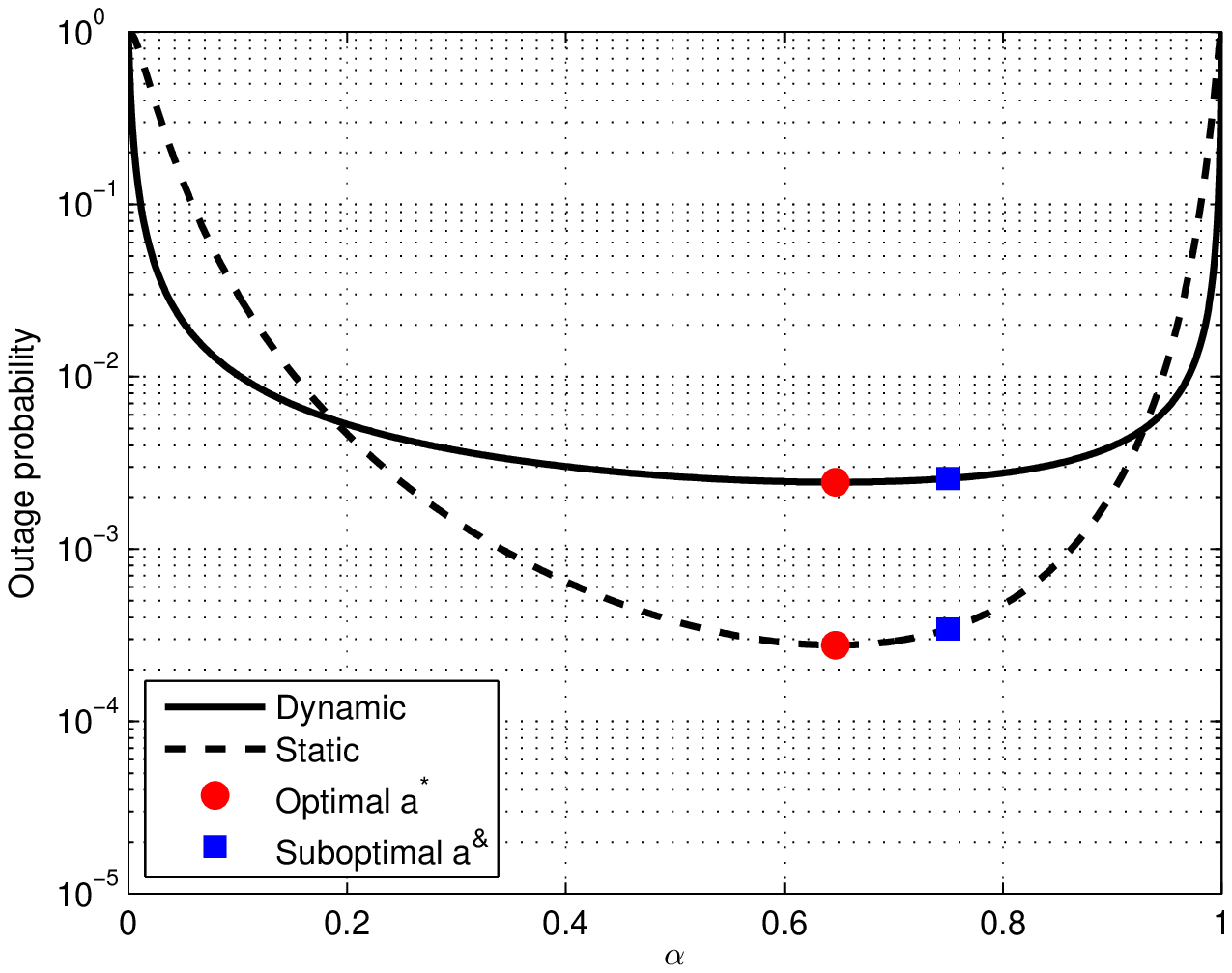}
        \caption{Bulk selection scheme}
    \end{subfigure}%
~
    \begin{subfigure}[t]{0.5\textwidth}
        \centering
        \includegraphics[width=3.5in]{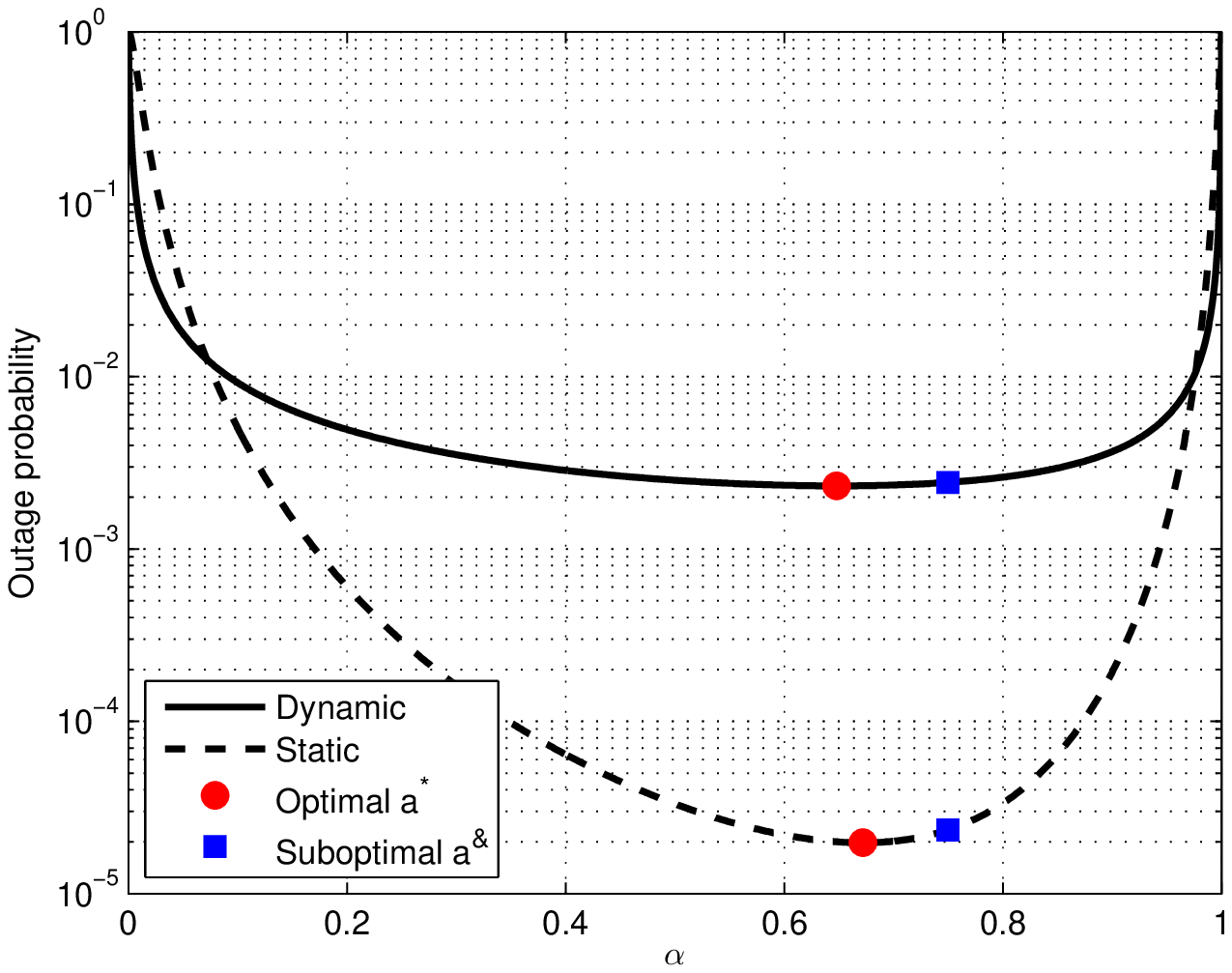}
        \caption{Per-subcarrier selection scheme}
    \end{subfigure}
    \caption{Outage probability of D2D communications vs. $\alpha$ for bulk and per-subcarrier selection schemes, given $N=K=4$.}
    \label{comparison_alpha}
\end{figure*}

\subsection{Verifications of outage performance optimization strategies}
To verify the effectiveness of the suboptimal solutions to the original optimization problem formulated in (\ref{diyigezuhewenti}), we carry out simulations to investigate the relation between $\alpha$ and outage probability in this subsection. Here we adopt the same simulation configurations as set in Section \ref{dsajkdhkj2dsdsd} and normalize $\bar{P}_S=\bar{P}_R=1$. Simulation results are shown in Fig. \ref{comparison_alpha}. Compared with the optimal and suboptimal power coordination factors $\alpha^*$ and $\alpha^\&$ (highlighted by red circles and blue squares), our proposed suboptimal solutions are close to the optimal solutions, which validate the feasibility of our proposed suboptimal algorithms for both dynamic and static power control cases. Therefore, we can employ these algorithms to efficiently coordinate transmit power among DUE transmitter and relays to yield a better outage performance.

\section{Conclusion}\label{c}
In this paper, we proposed an underlay OFDM D2D system, which is assisted by multiple full-duplex DF relays, and considered applying  multicarrier relay selections in this system. Meanwhile, power control mechanisms and performance optimizations were also taken into consideration in order to efficiently mitigate the interference from D2D communications to cellular communications. Then, we analyzed the outage performance of the proposed system. We obtained the single integral expressions of the outage probabilities when the dynamic power control mechanism was applied, and these expressions can be further simplified to closed forms when the static power control mechanism was utilized. After that, we studied the outage performance optimization problem by coordinating transmit power among DUE transmitter and relays. Due to the mathematical intractability of the original optimization problem, we proposed two alternative optimization problems, which are capable of providing suboptimal solutions for both dynamic and static power control cases. By the analytical and numerical results provided in this paper, we can have an insight into the relay-assisted OFDM D2D system, and understand its characteristics in most aspects thoroughly. Moreover, as a number of comparisons are given, this paper can also provide a guideline for implementing D2D communications and the relevant technologies in next generation networks.

\appendices

\section{Proof of Equivalence of Optimization Problems Formulated in (\ref{diyigezuhewenti}) and (\ref{hengchangjuns2})}\label{tilongbas}
According to the theorem of Lebesgue-Stieltjes integration, for an arbitrary random variable $X$ with CDF $F_X(x)$, we can derive its expectation as follows \cite{hajek2009notes}:
\begin{equation}\small\label{dsa3413521}
\mathbb{E}\{X\}=\int_{0}^{\infty}(1-F_X(x))\mathrm{d}x-\int_{-\infty}^{0} F_X(x)\mathrm{d}x.
\end{equation} 
In our case, because $X$ is the end-to-end SIR and $F_X(x)$ is the outage probability, it is obvious that $F_X(x)=0$  for $x<0$, and thus (\ref{dsa3413521}) can be simplified to 
\begin{equation}\small\label{dsa18cjwshxyhw}
\mathbb{E}\{X\}=\int_{0}^{\infty}(1-F_X(x))\mathrm{d}x.
\end{equation}
Considering $F_X(x)$ is a monotone increasing function of $x$, the relation given in (\ref{dsa18cjwshxyhw}) validates the equivalence between $\underset{\alpha}{\min}\{P_{out}(s)\}$ and $\underset{\alpha}{\max} \{\mathbb{E}\{\Gamma_{SR_nD}(k)\}\}$, for $0<\alpha<1$.

\section{Proof of Quasi-Concavity of the Optimization Problem Formulated in (\ref{qiuda23psjnde2122})}\label{dongtaizhengm}
To prove the quasi-concavity of the formulated problem in (\ref{qiuda23psjnde2122}), we first propose a lemma as follows:
\begin{lemma}\label{liemadsajhkd2}
Given a bounded, continuous and real piecewise function
\begin{equation}\small
f(x)=\begin{cases}
f_1(x),~~~~x\in(x_{\min},x_c)\\
f_2(x),~~~~x\in[x_c,x_{\max})
\end{cases},
\end{equation}
if $f_1(x)$ is a monotone increasing function of $x$, and $f_2(x)$ is a monotone decreasing function of $x$, $f(x)$ is a quasi-concave function of $x$ and the maximum $f(x)$ is achieved when $x=x_c$.
\end{lemma}
\begin{IEEEproof}
This lemma is straightforward to prove by elementary algebraic derivations and the definition of a quasi-concave function. Therefore, we omit a detailed proof here, and comprehensive analysis of the relation between quasi-concavity and monotonicity can be found in \cite{guerraggio2004origins}.
\end{IEEEproof}

As a result of \textit{Lemma \ref{liemadsajhkd2}}, we can transfer the exploration of quasi-concavity to the explorations of continuity and monotonicity. It is obvious from (\ref{omijia1}) and (\ref{omijia2}) that $\omega_1(\alpha)$ and $\omega_2(\alpha)$ are continuous over $\alpha\in(0,\varrho)$ and $\alpha\in[\varrho,1)$. Now, we can examine the continuity of both $\omega_1(\alpha)$ and $\omega_2(\alpha)$ at the boundary point $\alpha=\varrho$ by
\begin{equation}\small
\lim_{\alpha\rightarrow\varrho^{-}}\omega_i(\alpha)=\lim_{\alpha\rightarrow\varrho^{+}}\omega_i(\alpha),
\end{equation}
where $i\in\{1,2\}$. This relation indicates that both $\omega_1(\alpha)$ and $\omega_2(\alpha)$ are continuous at the boundary point $\alpha=\varrho$. Therefore, both $\omega_1(\alpha)$ and $\omega_2(\alpha)$ are continuous over the entire domain of definition $\alpha\in(0,1)$.

To investigate the monotonicity of $\omega_1(\alpha)$ and $\omega_2(\alpha)$, we propose another two lemmas as follows:
\begin{lemma}\label{hansukeeee}
$f(x)=Ax\left(1-e^{-\frac{B}{x}}\right)$, where $A$ and $B$ are bounded and positive constants, is a monotone increasing function of $x$, $\forall~x\in(0,1)$.
\end{lemma}
\begin{IEEEproof}
We derive the first and second order derivatives of $f(x)$ with respect to $x$ as follows:
\begin{equation}\small
f'(x)=\frac{\mathrm{d}f(x)}{\mathrm{d}x}=A\left(1-\frac{B+x}{x}e^{-\frac{B}{x}}\right)
\end{equation}
and 
\begin{equation}\label{sdkasjkd2erjiedao}\small
f''(x)=\frac{\mathrm{d}^2f(x)}{\mathrm{d}x^2}=-\frac{AB^2}{x^3}e^{-\frac{B}{x}}<0.
\end{equation}
From (\ref{sdkasjkd2erjiedao}), we know that $f'(x)$ is a monotone decreasing function of $x$ and therefore
\begin{equation}\small
\min_{x} \{f'(x)\}>\lim_{x\rightarrow 1}f'(x)=Ae^{-B}\left(e^{B}-B-1\right)>0.
\end{equation}
Because the first order derivative $f'(x)>0$, $\forall~x\in(0,1)$, $f(x)$ is a monotone increasing function of $x$.
\end{IEEEproof}
\begin{lemma}\label{hansukeeee2}
$t(x)=A(1-x)\left(1-e^{-\frac{B}{1-x}}\right)$, where $A$ and $B$ are bounded and positive constants, is a monotone decreasing function of $x$, $\forall~x\in(0,1)$.
\end{lemma}
\begin{IEEEproof}
We can express $t(x)=f(1-x)$. By \textit{Lemma \ref{hansukeeee}}, $f(x)$ is a monotone increasing function of $x$, and $f'(x)>0$, $\forall~x\in(0,1)$. Then, we can obtain the first order derivative of $t(x)$ with respect to $x$ by
\begin{equation}\small
t'(x)=\frac{\mathrm{d}t(x)}{\mathrm{d}x}=\frac{\mathrm{d}f(1-x)}{\mathrm{d}x}=-f'(x)<0,
\end{equation}
and therefore $t(x)=f(1-x)$ is a monotone decreasing function of $x$, $\forall~x\in(0,1)$.
\end{IEEEproof}
According to \textit{Lemma \ref{hansukeeee}} and \textit{Lemma \ref{hansukeeee2}}, we can easily see that $\omega_1(\alpha)$ and $\omega_2(\alpha)$ are monotone increasing functions of $\alpha$ when $0<\alpha<\varrho$ and monotone decreasing functions of $\alpha$ when $\varrho\leq\alpha<1$. As a result of \textit{Lemma \ref{liemadsajhkd2}}, we prove that $\omega_1(\alpha)$ and $\omega_2(\alpha)$ are quasi-concave functions of $x$, $\forall~x\in(0,1)$, and so as the quasi-concavity of the formulated problem in (\ref{qiuda23psjnde2122}).

\section{Proof of Quasi-Concavity of the Optimization Problem Formulated in (\ref{musat})}\label{sajdksahdakwu}
It can be easily seen that $\gamma_1(\alpha)$ is a bounded, continuous and monotone increasing function of $\alpha$, while $\gamma_2(\alpha)$ is a bounded, continuous and monotone decreasing function of $\alpha$. To prove $\gamma(\alpha)$ to be a quasi-concave function of $\alpha$, we first divide the formulated problem into three cases:
\begin{enumerate}
\item Case 1: $\gamma_1(\alpha)>\gamma_2(\alpha)$, $\forall~\alpha\in(0,1)$.
\item Case 2: $\gamma_1(\alpha)<\gamma_2(\alpha)$, $\forall~\alpha\in(0,1)$.
\item Case 3: $\gamma_1(\alpha)<\gamma_2(\alpha)$, for $\alpha\in(0,\epsilon)$, and $\gamma_1(\alpha)>\gamma_2(\alpha)$, $\forall~\alpha\in(\epsilon,1)$, where $\epsilon$ is a critical point in which $\gamma_1(\epsilon)=\gamma_2(\epsilon)$.
\end{enumerate}
For Case 1 and Case 2, it is straightforward that $\gamma(\alpha)=\gamma_1(\alpha)$ and $\gamma(\alpha)=\gamma_2(\alpha)$. Because of the monotonicity of $\gamma_1(\alpha)$ and $\gamma_2(\alpha)$, it is easy to derive the relation infra for both cases
\begin{equation}\small
\begin{split}
&\forall~\alpha_1,\alpha_2\in(0,1)~\mathrm{and}~\lambda\in(0,1),\\
&~~~~\exists~\gamma(\lambda\alpha_1+(1-\lambda)\gamma_2)\geq \min\left\lbrace\gamma(\alpha_1),\gamma(\gamma_2)\right\rbrace.
\end{split}
\end{equation}
Hence, according to the definition of a quasi-concave function \cite{concave}, we have proved $\gamma(\alpha)$ to be quasi-concave for Case 1 and Case 2. For Case 3, we suppose $0<\alpha_1<\alpha_2<1$ without losing generality and further divide Case 3 into another three sub-cases as follows:
\begin{enumerate}
\item Case 3-1: $0<\alpha_1<\alpha_2<\epsilon$
\item Case 3-2: $\epsilon<\alpha_1<\alpha_2<1$
\item Case 3-3: $0<\alpha_1<\epsilon<\alpha_2<1$
\end{enumerate}
Case 3-1 and Case 3-2 are simply special cases of Case 2 and Case 1, respectively. Therefore, the quasi-concavity of Case 3-1 and Case 3-2 can be proved in a similar manner as above. For Case 3-3, we need to discuss $\lambda$ in the range of $\left(0,\frac{\alpha_2-\epsilon}{\alpha_2-\alpha_1}\right)$ and $\left(\frac{\alpha_2-\epsilon}{\alpha_2-\alpha_1},1\right)$, respectively. When $\lambda\in\left(0,\frac{\alpha_2-\epsilon}{\alpha_2-\alpha_1}\right)$, we can derive
\begin{equation}\label{tako11}\small
\begin{split}
&\gamma(\lambda\alpha_1+(1-\lambda)\alpha_2)=\gamma_1(\lambda\alpha_1+(1-\lambda)\alpha_2)\geq \gamma_1(\alpha_1)=\gamma(\alpha_1)\geq \min\{\gamma(\alpha_1),\gamma(\alpha_2)\}.
\end{split}
\end{equation}
Similarly, when $\lambda\in\left(\frac{\alpha_2-\epsilon}{\alpha_2-\alpha_1},1\right)$, we can derive
\begin{equation}\label{tako12}\small
\begin{split}
&\gamma(\lambda\alpha_1+(1-\lambda)\alpha_2)=\gamma_2(\lambda\alpha_1+(1-\lambda)\alpha_2)\geq \gamma_2(\alpha_2)=\gamma(\alpha_2)\geq \min\{\gamma(\alpha_1),\gamma(\alpha_2)\}.
\end{split}
\end{equation}
By (\ref{tako11}) and (\ref{tako12}), we have proved $\gamma(\alpha)$ to be a quasi-concave function of $\alpha$ for Case 3. Now, we have proved $\gamma(\alpha)$ to be a quasi-concave function of $\alpha$ for all cases and thus the quasi-concavity of the formulated optimization problem.

\bibliographystyle{IEEEtran}
\bibliography{bib}

\end{document}